\documentstyle[psfig]{mn}

\def\etal{{\it et al.\ }}
\def\eg{{\it e.g.\ }}

\def\ie{{\it i.e.\ }}

\def\spose#1{\hbox to 0pt{#1\hss}}
\def\approxlt{\mathrel{\spose{\lower 3pt\hbox{$\sim$}}
	\raise 2.0pt\hbox{$<$}}}
\def\approxgt{\mathrel{\spose{\lower 3pt\hbox{$\sim$}}
	\raise 2.0pt\hbox{$>$}}}
\def\approxpropto{\mathrel{\spose{\lower 3pt\hbox{$\sim$}}
	\raise 2.0pt\hbox{$\propto$}}}
\mathchardef\twiddle="2218

\def\multleft#1{\hbox to size{\vbox {\halign {\lft{##}\cr #1}}\hfill}\par}
\def\multright#1{\hbox to size{\vbox {\halign {\rt{##}\cr #1}}\hfill}\par}
\def\Mdot{\hbox{$\dot M$}}

\def\today{\ifcase\month\or January\or February\or March\or April\or May\or
      June\or July\or August\or September\or October\or November\or December\fi
      \space\number\day, \number\year}
\def\<{\thinspace}

\def\apc{\rm atom cm$^{-2}$}

\def\cm{{\rm\thinspace cm}}
\def\erg{{\rm\thinspace erg}}

\def\keV{{\rm\thinspace keV}}

\def\km{{\rm\thinspace km}}
\def\kpc{{\rm\thinspace kpc}}

\def\Mpc{{\rm\thinspace Mpc}}
\def\Msun{\hbox{$\rm\thinspace M_{\odot}$}}

\def\s{{\rm\thinspace s}}
\def\yr{{\rm\thinspace yr}}

\def\ergpcmsqps{\hbox{$\erg\cm^{-2}\s^{-1}\,$}}

\def\ergps{\hbox{$\erg\s^{-1}\,$}}

\def\kmps{\hbox{$\km\s^{-1}\,$}}

\def\Msunpyr{\hbox{$\Msun\yr^{-1}\,$}}

\def\kmpspMpc{\hbox{$\kmps\Mpc^{-1}$}}

\title[ASCA and ROSAT observations of distant massive cooling flows]
{ASCA and ROSAT observations of distant massive cooling flows}
\author[S.W. Allen \etal ]
{\parbox[]{6.in} {S.W. Allen$^1$, A.C. Fabian$^1$, A.C. Edge$^1$,
M.W. Bautz$^2$, A. Furuzawa$^3$, Y. Tawara$^3$. \\
\footnotesize
1. Institute of Astronomy, Madingley Road, Cambridge CB3 OHA\\
2. Center for Space Research, MIT, Cambridge MA, USA\\
3. Department of Astrophysics, Nagoya University, Chikusa-ku, Furo-cho, Nagoya 464, Japan}}

\begin{document}

\date{Accepted for publication 1996 June}

\maketitle

\begin{abstract}

We present the results from a detailed ASCA/ROSAT X-ray study of three 
distant, massive cooling flows;  Zwicky 3146 ($z=0.291$), Abell 1835
($z=0.252$) and E1455+223 (Zwicky 7160;  $z=0.258$). Using multiphase 
models fitted to the ASCA spectra, we determine values for the temperature,
metallicity, luminosity and cooling  rates in the
clusters. These results are combined with deprojection analyses
of the ROSAT images to provide detailed constraints on the mass
distributions in the systems, and on the properties of their cooling flows. 
The spectral and imaging data identify these clusters as the three 
most-massive cooling flows known, with mass deposition rates of 
$\sim 1400, 2300$, and 1500 \Msunpyr~respectively. 
We highlight the need for multiphase models to consistently model
the spectral and imaging X-ray data, and discuss the importance 
of using these models in X-ray determinations of the cluster masses. 
We also present results from an extensive optical study of the clusters 
and report the discovery of a gravitational arc in Abell 1835. 
The lensing data provide an independent constraint on the
distribution of mass in the cluster, in good
agreement with the results from the multiphase (although not single-phase)
X-ray analysis.
We present measurements of the galaxy distributions in Abell 1835 and E1455+223 
and relate these to the distributions of the total mass. 
The ASCA spectra place firm constraints on the column density of intrinsic
X-ray absorbing material in the clusters. Abell 1835 and E1455+223 exhibit large 
intrinsic column densities ($N_{\rm H} \sim  3-4 \times 10^{21}$ \apc) 
associated with their cooling flows. These clusters also exhibit 
significant amounts of reddening in the optical spectra of their central galaxies. The data for
Zwicky 3146 indicate lower lower levels of X-ray absorption and reddening.  
All three clusters exhibit excellent alignment between the position angles of their 
X-ray emission and the optical emission from their dominant cluster galaxies.

\end{abstract}


\section{Introduction}

X-ray observations of clusters of galaxies show that in the
central regions of most clusters the cooling time of the IntraCluster Medium
(ICM) is significantly less than the Hubble time (\eg Edge, Stewart \& Fabian
1992).  The observed cooling, which takes place primarily through the
emission of X-rays, leads to a slow net inflow of material towards the cluster 
centre; a process known as a {\it cooling flow} (Fabian 1994). 

Within the idealized model of an {\it homogeneous} cooling flow, all of the cooling
gas flows to the centre of the cluster where it is deposited, having
radiated away its thermal energy. 
However, observations of cooling
flows show that the simple homogeneous model is not correct. 
Although the X-ray surface brightness profiles of clusters with  cooling flows
are substantially more sharply-peaked than those of non cooling-flow
systems, the emission is not as sharply-peaked as it would be in the case of  homogeneous
flows. Rather than all of the cooling gas flowing to the very
centres of the clusters, the X-ray data show that
gas is `cooling out' throughout the central few tens to hundreds of
kpc. Typically the cooled material is deposited with ${\dot M}(r) \propto
r$, where ${\dot M}(r)$ is the integrated mass deposition rate within radius
$r$. The X-ray data firmly require that cooling flows
are {\it inhomogeneous} with a range of density and temperature phases at all
radii. 

Spatially resolved X-ray spectroscopy of clusters also confirms the the presence of
distributed cool (and rapidly cooling) gas in cooling flows. The spatial distribution 
and luminosity of the cool components determined from the spectral data are well-matched  
to values inferred from the X-ray images (\eg Allen \etal 1993, Allen, Fabian \& Kneib
1996; Fabian \etal 1996; Allen \& Fabian 1996).  
For a detailed  review of the theory and observations of cooling flows see
Fabian (1994).

In this paper we present observations, made with the  ASCA and ROSAT  X-ray
astronomy satellites, of three exceptionally luminous cluster cooling
flows. Two of the systems, Zwicky 3146 ($z=0.291$) and Abell 1835 ($z=0.252$)  were identified as 
X-ray luminous clusters during optical-follow up studies to 
the ROSAT All-Sky Survey (Allen \etal 1992a).  The third system, 
E1455+223 (or Zwicky 7160; $z=0.258$) was identified 
by Mason
\etal (1981) in a follow-up to X-ray observations made with 
Einstein Observatory. All three clusters are included in the ROSAT
Brightest Cluster Sample (Ebeling \etal 1996a).

Combining the high spectral resolution ASCA data with ROSAT images 
we present consistent determinations of the temperatures, metallicities, luminosities 
and cooling rates in the clusters. 
The data for Zwicky 3146, Abell 1835 and E1455+223 identify them as 
the three largest cluster cooling flows known to date. We constrain the level 
of intrinsic X-ray absorption in the cooling flows and relate the results to
measurements of intrinsic reddening in
the Central Cluster Galaxies (CCGs) of the systems. 
Results on the distributions of X-ray gas, galaxies, and the total gravitating
matter in the clusters are reported.

The structure of this paper is as follows. In Section 2 we summarize  the
observations. In Section 3 we present the X-ray and optical imaging data 
and discuss the morphological relationships between the
clusters and their CCGs. In Section 4 we discuss
the spectral analysis of the X-ray data. In Section 5 present the results from the 
deprojection analyses of the ROSAT images. Section 6 discusses the 
optical properties of the clusters. In Section 7 we discuss some of the more 
important results in detail, and in Section 8 summarize our conclusions.  
Throughout this paper, we assume $H_0$=50
\kmpspMpc, $\Omega = 1$ and $\Lambda = 0$.

\section{Observations}

\begin{table*}
\vskip 0.2truein
\begin{center}
\caption{Observation summary}
\vskip 0.2truein
\begin{tabular}{ c c c c c c c }

\hline                                                                                                                     
 Cluster      & ~ &   Instrument       & ~ &       Observation Date      & ~ &  Exposure (ks)   \\                                
&&&&&& \\                                                                                                                   
Zwicky 3146   & ~ &   ASCA SIS0        & ~ &       1993 May 18           & ~ &          26.3              \\
              & ~ &   ASCA SIS1        & ~ &          " "                & ~ &          31.6              \\
              & ~ &   ASCA GIS2        & ~ &          " "                & ~ &          30.2              \\
              & ~ &   ASCA GIS3        & ~ &          " "                & ~ &          30.2              \\
              & ~ &   ROSAT HRI \#1    & ~ &       1992 Nov 27           & ~ &          15.2              \\
              & ~ &   ROSAT HRI \#2    & ~ &       1993 May 17           & ~ &          10.8              \\
              & ~ &   ROSAT PSPC       & ~ &       1993 Nov 13           & ~ &          8.62              \\
              & ~ &   ESO 3.6m (R)     & ~ &       1992 Nov 11           & ~ &          0.12              \\
&&&&&& \\                                                                                                                    
Abell 1835    & ~ &   ASCA SIS0 \#1    & ~ &       1994 Jul 20           & ~ &          18.2              \\
              & ~ &   ASCA SIS1 \#1    & ~ &         " "                 & ~ &          17.2              \\
              & ~ &   ASCA GIS2 \#1    & ~ &         " "                 & ~ &          13.0              \\
              & ~ &   ASCA GIS3 \#1    & ~ &         " "                 & ~ &          13.0              \\
              & ~ &   ASCA SIS0 \#2    & ~ &       1994 Jul 21           & ~ &          8.51              \\
              & ~ &   ASCA SIS1 \#2    & ~ &         " "                 & ~ &          8.22              \\
              & ~ &   ASCA GIS2 \#2    & ~ &         " "                 & ~ &          6.55              \\
              & ~ &   ASCA GIS3 \#2    & ~ &         " "                 & ~ &          6.56              \\
              & ~ &   ROSAT HRI        & ~ &       1993 Jan 22           & ~ &          2.85              \\
              & ~ &   ROSAT PSPC       & ~ &       1993 Jul 03           & ~ &          6.18              \\
              & ~ &   Hale 5m (Gunn i) & ~ &       1994 Jun 09            & ~ &          1.00              \\
              & ~ &   Hale 5m (KC B)   & ~ &       1994 Jun 10           & ~ &          0.50              \\
              & ~ &   Hale 5m (KC U)   & ~ &       1994 Jun 09            & ~ &          3.00               \\
&&&&&& \\                                                                                                                  
E1455+223     & ~ &   ASCA SIS0        & ~ &       1994 Jul 18           & ~ &          30.5              \\
              & ~ &   ASCA SIS1        & ~ &          " "                & ~ &          28.3              \\
              & ~ &   ASCA GIS2        & ~ &          " "                & ~ &          18.3              \\
              & ~ &   ASCA GIS3        & ~ &          " "                & ~ &          18.3              \\
              & ~ &   ROSAT HRI \#1    & ~ &       1992  Jan 11          & ~ &          4.09              \\
              & ~ &   ROSAT HRI \#2    & ~ &       1993  Jan 20          & ~ &          4.23              \\
              & ~ &   ROSAT HRI \#3    & ~ &       1994  Jul 07          & ~ &          6.57              \\
              & ~ &   Hale 5m (Gunn i) & ~ &       1994  Jun 10          & ~ &          0.50              \\
              & ~ &   Hale 5m (KC B)   & ~ &       1994  Jun 10          & ~ &          0.60              \\
              & ~ &   Hale 5m (KC U)   & ~ &       1994  Jun 10          & ~ &          3.00              \\
\hline                                                                                                                     
&&&&&& \\

\end{tabular}
\end{center}
\parbox {7in}
{Notes: Exposure times are effective exposures after all 
cleaning and correction procedures have been carried out. For Abell 1835 the 
ASCA observations were carried out in 2 parts, yielding total exposures of 26.7, 
25.4, 19.6 and 19.6 ks for the S0, S1, G2 and G3 detectors, respectively. For Zwicky 3146,
two separate ROSAT HRI observations were made, yielding a total exposure of 26.0
ks. For E1455+223 three HRI observations  were carried out providing a  total
exposure of 14.9 ks. }
\end{table*}

\begin{table*}
\vskip 0.2truein
\begin{center}
\caption{Target summary}
\vskip 0.2truein
\begin{tabular}{ c c c c c c c c c c c c c c c c }
\hline                                                                                                                                                                                                                                                                                          
\multicolumn{1}{c}{} &
\multicolumn{1}{c}{} &
\multicolumn{1}{c}{} &
\multicolumn{2}{c}{OPTICAL (J2000.)} &
\multicolumn{2}{c}{X-RAY (J2000.)} &
\multicolumn{1}{c}{} &
\multicolumn{1}{c}{} \\
Cluster     & ~ &  $z$   &  R.A.                              & Dec.                &  R.A.                              & Dec.                 & $F_X$  & $L_X$  \\
&&&&&&&& \\                                                                                                                                
Zwicky 3146 & ~ & 0.2906 & $10^{\rm h}23^{\rm m}39.6^{\rm s}$ & $04^{\circ}11'10''$ & $10^{\rm h}23^{\rm m}39.8^{\rm s}$ & $04^{\circ}11'11''$  & 6.6    & 2.8    \\
Abell 1835  & ~ & 0.2523 & $14^{\rm h}01^{\rm m}02.0^{\rm s}$ & $02^{\circ}52'42''$ & $14^{\rm h}01^{\rm m}01.9^{\rm s}$ & $02^{\circ}52'43''$  & 12.3   & 3.8    \\
E1455+223  & ~ & 0.2578 & $14^{\rm h}57^{\rm m}15.1^{\rm s}$ & $22^{\circ}20'31''$ & $14^{\rm h}57^{\rm m}15.0^{\rm s}$ & $22^{\circ}20'36''$  & 3.7    & 1.3    \\
\hline                                                                                                                                                                                                                                                                                          
&&&&&&&& \\                                                                                                                                

\end{tabular}
\end{center}

\parbox {7in}
{Notes: Redshifts and CCG coordinates (J2000) from
Allen \etal (1992).  X-ray coordinates denote the position of the X-ray peak determined from the 
HRI data. X-ray fluxes ($F_X$) in units of $10^{-12}$ \ergpcmsqps and luminosities ($L_X$) 
in $10^{45}$ \ergps~are determined from the ASCA (S0) data. Fluxes are quoted in the 2-10 keV band of the observer. Luminosities
are absorption-corrected and are quoted in the 2-10 keV rest-frame of the object.}

\end{table*}

The details of the observations are summarised in Table 1.   Exposure times for
the ASCA data sets are effective exposures after standard data screening and 
cleaning procedures have been applied  (Day \etal 1995). Hashed numbers 
following an instrument name indicate that those observations were carried out
on more than one date. The ASCA observations of Abell 1835 were made in two
parts, on consecutive days in 1994 July. 
The ROSAT HRI observations of Zwicky 3146 were carried out on 2 dates, in 1992 
November and 1993 
May,  giving a
total exposure of 26.0 ks. The HRI observations of E1455+223 were carried out
in 3 parts, in 1992 January -- 1994 July,  with a total exposure time of 14.9 ks.

The ASCA observation of Zwicky 3146 was carried out  in 1993 May during the PV
stage of the  mission. The SIS detectors were used in 2-CCD  mode with
the target positioned approximately at the boundary between chips 1 and 2 in
SIS0  (the nominal pointing position in 2-CCD mode  during the PV phase). The
observations of Abell 1835 and E1455+223 were carried out during the AO-1 stage
of the ASCA  program. These clusters were also observed in 2-CCD mode, but with
the targets  positioned more centrally in chip 1 of SIS0. For a detailed
discussion of ASCA observing modes and instrument configurations see Day \etal
(1995). Reduction of the ASCA data was carried out using the FTOOLS package. 
Standard selection and screening criteria were applied (Day \etal 1995).
The ROSAT data were analysed using the STARLINK ASTERIX package.

The optical observations of Zwicky 3146 was carried out with the 3.6m 
telescope at the European Southern Observatory (ESO), La Silla, Chile.  The ESO
Faint Object Spectrograph Camera  was used with the TEK $512 \times 512$ CCD
(pixel scale 0.61 arcsec).   An exposure of 120 sec in the R band was made  in
seeing of $\sim 1.5$ arcsec.  The optical observations of Abell 1835 and
E1455+223 were carried out with the  5m Hale Telescope, Palomar, 
as part of a follow up study (Edge \etal 1996) of
the most X-ray luminous clusters in the ROSAT Brightest Cluster Sample.
The COSMIC instrument and TEK $2048 \times 2048$ chip (pixel scale 0.28
arcsec) were used.
Exposures of 1000 and 500 s were made in Gunn i, 500s and 600s in  KC B, and 3000s in KC U, 
for Abell 1835 and E1455+223  respectively. The seeing was $\sim 1.1$ arcsec. 
The optical data were reduced and analysed in IRAF.

Figs. 1-3 show the optical images of the clusters, with the 
ROSAT HRI X-ray contours overlaid. (Details of the smoothing algorithms are
given in the figure captions.)

\begin{figure*}
\vskip12.5cm
\caption{ The ESO 3.6m R band image of Zwicky 3146 with the ROSAT  HRI X-ray
contours overlaid. The optical data have a pixel scale of $0.61$ arcsec and
were taken in $\sim 1.5$ arcsec seeing. The X-ray image has a pixel size of  $2
\times 2$ arcsec$^2$ and has been adaptively smoothed (Ebeling, White \& Rangarajan 1996) 
to give $\geq 36$ count smoothing element$^{-1}$. Contours are drawn at eight evenly-spaced 
logarithmic intervals between 0.89 and 22.38 ct pixel$^{-1}$.
}
\end{figure*}

\begin{figure*}
\vskip12.5cm
\caption{ The Hale 5m Gunn i image of Abell 1835 with the ROSAT  HRI X-ray
contours overlaid. The optical data have a pixel scale of $0.28$ arcsec and
were taken in $\sim 1.1$ arcsec seeing. The X-ray image has a pixel size
of  $4 \times 4$ arcsec$^2$ and has been adaptively smoothed  to give
$\geq 16$ count smoothing element$^{-1}$. Contours are drawn at six evenly-spaced logarithmic
intervals between 0.79 and 7.94 ct pixel$^{-1}$.
}
\end{figure*}

\begin{figure*}
\vskip12.5cm
\caption{ The Hale 5m Gunn i image of E1455+223 with the ROSAT  HRI X-ray
contours overlaid. The optical data have a pixel scale of $0.28$ arcsec and
were taken in $\sim 1.1$ arcsec seeing. The X-ray image has a pixel size of  $2
\times 2$ arcsec$^2$ and has been adaptively smoothed  to give $\geq 25$ count
smoothing element$^{-1}$. Contours are drawn at nine evenly-spaced logarithmic
intervals between 0.21 and 8.32 ct pixel$^{-1}$.
}
\end{figure*}

\section{Morphology Analysis}

\begin{table*}
\vskip 0.2truein
\begin{center}
\caption{Isophote Analysis}
\vskip 0.2truein
\begin{tabular}{ c c c c c c c c c c c c c }
\hline                                                                                                                                                                                                                                                                                          
\multicolumn{1}{c}{Cluster} &
\multicolumn{1}{c}{} &
\multicolumn{4}{c}{OPTICAL} &
\multicolumn{1}{c}{} &
\multicolumn{4}{c}{X-RAY} \\                            
            & ~ & pixel                & range     &  ellipticity     &   P.A.         & ~~ &    pixel         & range    &  ellipticity   &  P.A.           \\  
&&&&&&&&& \\                                                                               
Zwicky 3146 & ~ & $0.61 \times  0.61$  & $1.2-3.7$ & $0.32 \pm 0.01$  & $125 \pm 1$    & ~~ & $8.0 \times 8.0$ & $16-24$  & $0.16 \pm 0.04$  & $127 \pm 8$    \\
Abell 1835  & ~ & $0.57 \times  0.57$  & $1.1-4.5$ & $0.21 \pm 0.03$  & $145 \pm 4$    & ~~ & $8.0 \times 8.0$ & $16-56$  & $0.20 \pm 0.07$  & $163 \pm 16$   \\
E1455+223   & ~ & $0.57 \times  0.57$  & $1.1-4.5$ & $0.16 \pm 0.01$  & $35  \pm 2$    & ~~ & $8.0 \times 8.0$ & $16-48$  & $0.20 \pm 0.05$  & $40  \pm 7$    \\
\hline                                                                                                                                                                                                                                                                                          
&&&&&&&&& \\

\end{tabular}
\end{center}

\parbox {7in}
{Notes: A summary of the results from the isophote analysis of the  optical
(CCG) and X-ray (cluster) data. Columns (2) and (6) give the pixel sizes in
arcsec of the re-binned optical and X-ray images used in the analyses. Columns
(3) and  (7) list the range (in arcsec) of semi-major axes analysed.  Columns 
(4) and (8)
give the mean ellipticities (defined as $1-b/a$ where $b$ and $a$ are  the
semi-minor and semi-major axes respectively) of the optical and X-ray data in
the regions analysed. Columns (5) and (9) list the mean position angles (PA) 
in degrees in these same regions.} 
\end{table*}

The images presented in Figs. 1-3,  and the optical and X-ray co-ordinates listed in Table 2, 
demonstrate  excellent agreement between the positions of the CCGs and the 
positions of the peaks of the 
X-ray emission from the clusters.  [Errors of $\approxlt 5$ arcsec may be
associated with the aspect solutions of the HRI data. The astrometry of the
optical data is accurate to within 1 arcsec.]

The ellipticities and position angles of the X-ray emission from the  clusters
and the optical emission from the CCGs have been examined using the ELLIPSE
isophote-analysis routines in IRAF.  The images were re-binned to a suitable
pixel size ($8 \times 8$ arcsec$^2$ for the X-ray data and $0.57 \times 0.57$
arcsec$^2$ for the optical images) and were  modelled with elliptical isophotes
(Jedrzejewski 1987). The ellipticities, position angles and centroids of the
isophotes were  free parameters in the fits. The results are summarized in 
Table 3 where we list the mean ellipticities and position angles  over the
range of semi-major axes studied. We find  excellent agreement between the
position angles of the optical (CCG) and X-ray (cluster) isophotes. 

The agreement between the  position angles of the CCG and cluster isophotes,
and the coincidence of the CCG coordinates and the peaks of the cluster X-ray
emission, are similar to the results from studies of other  large cooling-flow
clusters at lower redshifts (White \etal 1994;  Allen \etal 1995; Allen \etal 1996).

\section{Spectral Analysis of the ASCA data}

\subsection{Method of Analysis}

\begin{table}
\vskip 0.2truein
\begin{center}
\caption{Regions included in the spectral analysis}
\vskip 0.2truein
\begin{tabular}{ c c c c c }
&&&& \\                                                     

\hline                                                                                                                                                                                                                                                                                          
 Cluster    & ~ &  Detector & ~ &   Radius (arcmin/kpc)   \\
&&&& \\
Zwicky 3146 & ~ &   SIS0     & ~ &        5.0/1620         \\
            & ~ &   SIS1     & ~ &        4.0/1230         \\
            & ~ &   SIS2     & ~ &        6.0/1940         \\
            & ~ &   SIS3     & ~ &        6.0/1940         \\
&&&& \\
Abell 1835  & ~ &   SIS0     & ~ &        4.0/1190         \\
            & ~ &   SIS1     & ~ &        3.0/890          \\
            & ~ &   SIS2     & ~ &        6.0/1780         \\
            & ~ &   SIS3     & ~ &        6.0/1780         \\
&&&& \\
E1455+223   & ~ &   SIS0     & ~ &        3.8/1140         \\
            & ~ &   SIS1     & ~ &        2.8/840          \\
            & ~ &   SIS2     & ~ &        6.0/1800         \\
            & ~ &   SIS3     & ~ &        6.0/1800         \\
\hline
&&&& \\

\end{tabular}
\end{center}

\end{table}

SIS spectra were extracted from circular regions centred on the  positions of
the X-ray peaks. The radii of these regions were  selected to minimize the
number of chip boundaries crossed (and thereby minimize systematic
uncertainties introduced into the data by such crossings) whilst covering as
large a region of the clusters as possible.  The compromise of these
considerations lead to  the  choice of regions for spectral analysis listed in
Table 4. For Abell 1835 and E1455+223 the spectra were extracted from a single
chip in each SIS (chip 1 in SIS0 and chip 3 in SIS1).  For Zwicky 3146, which
is centred on the boundary between chips 1 and 2 in SIS0  (chips 3 and 0 in
SIS1), the data were extracted across the two chips  in circular  regions
bounded by the outer chip edges.   The GIS spectra used for the analysis of the
cluster properties  were extracted  from circular regions of radius 6 arcmin 
(corresponding to $\sim 2$ Mpc at the redshifts of the clusters) again centred
on the peak of the X-ray emission from the clusters.  

Background subtraction was carried out using the  `blank sky'
observations compiled during the performance verification stage of the ASCA
mission. (The blank sky observations are compiled from observations of  high
Galactic latitude fields free of  bright X-ray sources).   For X-ray sources
lying in directions of relatively low
Galactic column density, like the targets discussed in this paper,  the blank sky 
observations provide a reasonable
representation of the cosmic and instrumental backgrounds in the detectors over
the energy ranges of interest.

The modelling of the X-ray spectra has been carried out using the XSPEC 
spectral fitting package (version 8.50; Shafer \etal 1991).  For the SIS
analysis, the 1994 November 9 release of the response matrices from GSFC was
used.   Only those counts in pulse height analyser (PHA) channels corresponding
to  energies between 0.5 and 10  \keV~  were included in the fits (the  energies
between which the calibration of the SIS is best-understood). For the GIS
analysis, the 1995 March 6 release of the GSFC response matrices was used and
only data in the energy range $1  - 10$ \keV~were included in the fits.  All
spectra were grouped before fitting to ensure a minimum of 20  counts per PHA
channel,  thereby allowing $\chi^2$ statistics to be used. 

The X-ray emission from the clusters has been modelled using the plasma codes 
of Raymond \& Smith (1977; with updates  incorporated into XSPEC version 8.50)
and Kaastra \& Mewe (1993). The results for the two plasma codes show good
agreement. For clarity, only the results for the Raymond \& Smith (hereafter
RS) code will be presented in detail in this paper  although the conclusions
drawn may equally be applied to the analysis with the Kaastra \& Mewe  code.

We have modelled the ASCA spectra both by fitting the data from the individual detectors 
independently and by combining the data from all 4 detectors. The results from the 
fits to the individual detectors are summarised in Tables 5-7. When combining the data for the different detectors 
the temperature, metallicity and column density values were linked together. However, the 
normalizations of both the ambient cluster emission and the cooling flow components 
were allowed to vary independently, due to 
the different source extraction areas used and residual uncertainties in the flux calibration 
of the instruments. The results from the fits to the combined data sets are summarized in Table 8.

\subsection{The spectral models}

\begin{table*}
\vskip 0.2truein
\begin{center}
\caption{Spectral Analysis of the ASCA data for Zwicky 3146}
\vskip 0.2truein
\begin{tabular}{ c c c c c c c c c c c }
&&&&&&&&&& \\
\hline                                                                                                                                                         

           & ~ &   Parameters   & ~ &          S0            & ~~~ &     S1                 & ~~~ &         G2              & ~~~ &         G3             \\
&&&&&&&&&& \\
           & ~ &   $kT$         & ~ & $5.6^{+0.3}_{-0.3}$    & ~~~ & $5.6^{+0.3}_{-0.4}$    & ~~~ &  $6.2^{+0.6}_{-0.6}$    & ~~~ & $6.2^{+0.6}_{-0.6}$    \\ 
           & ~ &   $Z$          & ~ & $0.22^{+0.07}_{-0.06}$ & ~~~ & $0.30^{+0.09}_{-0.08}$ & ~~~ &  $0.19^{+0.11}_{-0.11}$ & ~~~ & $0.31^{+0.13}_{-0.11}$ \\ 
 MODEL A   & ~ &   $N_{\rm H}$  & ~ & $0.34$                 & ~~~ & $0.34$                 & ~~~ &  $0.34$                 & ~~~ & $ 0.34$               \\ 
           & ~ &   $\chi^2$/DOF & ~ & 243.7/235              & ~~~ & 256.1/207              & ~~~ &  182.3/209              & ~~~ & 173.9/214              \\ 
&&&&&&&&&& \\

           & ~ &   $kT$         & ~ & $5.9^{+0.7}_{-0.4}$    & ~~~ & $5.6^{+0.6}_{-0.5}$    & ~~~ &  $6.5^{+0.6}_{-0.7}$    & ~~~ & $6.3^{+0.8}_{-0.8}$    \\ 
           & ~ &   $Z$          & ~ & $0.22^{+0.07}_{-0.07}$ & ~~~ & $0.30^{+0.10}_{-0.08}$ & ~~~ &  $0.19^{+0.11}_{-0.11}$ & ~~~ & $0.32^{+0.12}_{-0.12}$ \\ 
 MODEL B   & ~ &   $N_{\rm H}$  & ~ & $0.16^{+0.15}_{-0.15}$ & ~~~ & $0.34^{+0.18}_{-0.17}$ & ~~~ &  $<0.39$                & ~~~ & $< 0.97$               \\ 
           & ~ &   $\chi^2$/DOF & ~ & 239.8/234              & ~~~ & 256.1/206              & ~~~ &  180.1/208              & ~~~ & 173.8/213              \\ 
&&&&&&&&&& \\

           & ~ &   $kT$         & ~ & $6.6^{+4.4}_{-1.0}$    & ~~~ & $5.6^{+2.3}_{-0.5}$    & ~~~ &  $11.2^{+8.2}_{-5.0}$   & ~~~ &  $12.3^{+4.4}_{-5.8}$    \\ 
           & ~ &   $Z$          & ~ & $0.23^{+0.08}_{-0.07}$ & ~~~ & $0.30^{+0.10}_{-0.09}$ & ~~~ &  $0.30^{+0.17}_{-0.19}$ & ~~~ &  $0.48^{+0.17}_{-0.19}$  \\ 
 MODEL C   & ~ &   $N_{\rm H}$  & ~ & $0.38^{+0.43}_{-0.30}$ & ~~~ & $0.42^{+0.58}_{-0.24}$ & ~~~ &  $<1.29$                & ~~~ &  $1.26^{+0.82}_{-0.96}$  \\ 
           & ~ &   ${\dot M}$   & ~ & $<2900$                & ~~~ & $<2800$                & ~~~ &  $<2290$                & ~~~ &  $2240^{+260}_{-1760}$   \\ 
           & ~ &   $\chi^2$/DOF & ~ & 238.5/233              & ~~~ & 256.0/205              & ~~~ &  178.0/207              & ~~~ &  170.2/212               \\ 
&&&&&&&&&& \\

           & ~ &   $kT$         & ~ & $6.6^{+2.4}_{-0.9}$    & ~~~ & $5.4^{+1.3}_{-0.5}$    & ~~~ & $11.5^{+6.9}_{-5.1}$    & ~~~ & $12.7^{+3.9}_{-6.1}$     \\ 
           & ~ &   $Z$          & ~ & $0.23^{+0.08}_{-0.08}$ & ~~~ & $0.31^{+0.08}_{-0.09}$ & ~~~ & $0.30^{+0.17}_{-0.17}$  & ~~~ & $0.49^{+0.15}_{-0.20}$   \\ 
 MODEL D   & ~ &   $N_{\rm H}$  & ~ & $0.34$                 & ~~~ & $0.34$                 & ~~~ & $0.34$                  & ~~~ & $0.34$                   \\ 
           & ~ &   ${\dot M}$   & ~ & $1150^{+1400}_{-850}$  & ~~~ & $<1100$                & ~~~ & $1740^{+570}_{-1340}$   & ~~~ & $2240^{+250}_{-1740}$    \\ 
           & ~ & $\Delta N_{\rm H}$  & ~ & $<1.1$                 & ~~~ & $U.C.$                 & ~~~ & $<2.44$                 & ~~~ & $<5.43$                  \\ 
           & ~ &   $\chi^2$/DOF & ~ & 238.6/233              & ~~~ & 255.5/205              & ~~~ & 177.9/207               & ~~~ & 170.2/212                \\ 
\hline                                                                                                                                                         
&&&&&&&&&& \\                                                                                                                                               

\end{tabular}
\end{center}

\parbox {7in}
{ Notes: The best-fit parameter values and 90 per cent  ($\Delta \chi^2 =
2.71$) confidence limits from the spectral analysis of the ASCA data for Zwicky 
3146.  Temperatures ($kT$) are in keV and metallicities ($Z$) are quoted  as a
fraction of the Solar value (Anders \& Grevesse 1989). Column densities
($N_{\rm H}$) are in units of $10^{21}$ atom cm$^{-2}$ and mass deposition
rates  (\Mdot) in \Msunpyr.   $\chi^2$ values and the number of degrees of
freedom (DOF) in the fits are given for the four spectral models discussed in
Section 4.2. }
\end{table*}

\begin{table*}
\vskip 0.2truein
\begin{center}
\caption{Spectral Analysis of the ASCA data for Abell 1835}
\vskip 0.2truein
\begin{tabular}{ c c c c c c c c c c c }
&&&&&&&&&& \\

\hline                                                                                                                                                         
           & ~ &   Parameters   & ~ &          S0            & ~~~ &     S1                 & ~~~ &         G2              & ~~~ &         G3             \\
&&&&&&&&&& \\
           & ~ &   $kT$         & ~ & $9.4^{+0.7}_{-0.6}$    & ~~~ & $9.4^{+1.1}_{-0.7}$    & ~~~ &  $6.5^{+0.7}_{-0.6}$    & ~~~ & $6.8^{+0.6}_{-0.7}$    \\ 
           & ~ &   $Z$          & ~ & $0.24^{+0.10}_{-0.09}$ & ~~~ & $0.26^{+0.12}_{-0.12}$ & ~~~ &  $0.26^{+0.13}_{-0.12}$ & ~~~ & $0.23^{+0.12}_{-0.11}$ \\ 
 MODEL A   & ~ &   $N_{\rm H}$  & ~ & $0.22$                 & ~~~ & $0.22$                 & ~~~ &  $0.22$                 & ~~~ & $0.22$               \\ 
           & ~ &   $\chi^2$/DOF & ~ & 416.7/369              & ~~~ & 363.5/298              & ~~~ &  200.0/216              & ~~~ & 199.5/261              \\ 
&&&&&&&&&& \\                                                                                                                                               

           & ~ &   $kT$         & ~ & $7.4^{+0.8}_{-0.6}$    & ~~~ & $7.1^{+0.9}_{-0.7}$    & ~~~ &  $6.7^{+0.8}_{-1.0}$    & ~~~ & $6.0^{+0.9}_{-0.7}$    \\ 
           & ~ &   $Z$          & ~ & $0.23^{+0.08}_{-0.07}$ & ~~~ & $0.26^{+0.09}_{-0.09}$ & ~~~ &  $0.26^{+0.14}_{-0.12}$ & ~~~ & $0.23^{+0.11}_{-0.10}$ \\ 
 MODEL B   & ~ &   $N_{\rm H}$  & ~ & $0.73^{+0.15}_{-0.14}$ & ~~~ & $0.81^{+0.19}_{-0.19}$ & ~~~ &  $<0.82$                & ~~~ & $1.04^{+0.75}_{-0.69}$ \\ 
           & ~ &   $\chi^2$/DOF & ~ & 379.2/368              & ~~~ & 334.4/297              & ~~~ &  199.9/215              & ~~~ & 195.7/260              \\ 
&&&&&&&&&& \\

           & ~ &   $kT$         & ~ & $8.8^{+6.6}_{-1.8}$    & ~~~& $7.2^{+5.0}_{-0.7}$    & ~~~ &  $12.1^{+3.3}_{-6.1}$   & ~~~ & $6.0^{+6.1}_{-0.7}$   \\ 
           & ~ &   $Z$          & ~ & $0.26^{+0.09}_{-0.08}$ & ~~~& $0.26^{+0.08}_{-0.09}$ & ~~~ &  $0.34^{+0.17}_{-0.17}$ & ~~~ & $0.23^{+0.13}_{-0.10}$  \\ 
MODEL C    & ~ &   $N_{\rm H}$  & ~ & $1.04^{+0.32}_{-0.40}$ & ~~~& $0.81^{+0.33}_{-0.19}$ & ~~~ &  $1.03^{+0.83}_{-0.64}$ & ~~~ & $1.05^{+1.07}_{-0.52}$  \\ 
           & ~ &   ${\dot M}$   & ~ & $<2600$                &~~~ & $<2600$                & ~~~ &  $<3500$                & ~~~ &  $<4400$                 \\ 
           & ~ &   $\chi^2$/DOF & ~ & 377.4/367              & ~~~& 334.4/296              & ~~~ &  198.3/214              & ~~~ &  195.7/259               \\ 
&&&&&&&&&& \\

           & ~ &   $kT$         & ~ & $9.7^{+3.5}_{-1.2}$    & ~~~ & $8.0^{+1.6}_{-1.2}$    & ~~~ & $9.5^{+7.5}_{-3.6}$    & ~~~ & $6.2^{+5.9}_{-0.9}$    \\ 
           & ~ &   $Z$          & ~ & $0.28^{+0.06}_{-0.08}$ & ~~~ & $0.28^{+0.11}_{-0.10}$ & ~~~ & $0.32^{+0.16}_{-0.16}$ & ~~~ & $0.24^{+0.13}_{-0.11}$   \\ 
 MODEL D   & ~ &   $N_{\rm H}$  & ~ & $0.22$                 & ~~~ & $0.22$                 & ~~~ & $0.22$                 & ~~~ & $0.22$                   \\ 
           & ~ &   ${\dot M}$   & ~ & $2000^{+550}_{-450}$   & ~~~ & $2000^{+1000}_{-600}$  & ~~~ & $<4300$                & ~~~ & $<4300$                  \\ 
           & ~ & $\Delta N_{\rm H}$  & ~ & $3.25^{+2.25}_{-0.85}$ & ~~~ & $6.68^{+5.50}_{-2.59}$ & ~~~ & U.C.                   & ~~~ & U.C.                     \\ 
           & ~ &   $\chi^2$/DOF & ~ & 372.9/367              & ~~~ & 327.2/296              & ~~~ & 198.9/214              & ~~~ & 195.5/259                \\ 
\hline                                                                                                                                                         
&&&&&&&&&& \\                                                                                                                                               

\end{tabular}
\end{center}

\parbox {7in}
{ Notes: The best-fit parameter values and 90 per cent ($\Delta \chi^2 = 2.71$)
confidence limits from the spectral analysis of the ASCA data for Abell 1835.
Details as for Table 5.}
\end{table*}

\begin{table*}
\vskip 0.2truein
\begin{center}
\caption{Spectral Analysis of the ASCA data for E1455+223}
\vskip 0.2truein
\begin{tabular}{ c c c c c c c c c c c }
&&&&&&&&&& \\

\hline                                                                                                                                                         
           & ~ &   Parameters   & ~ &          S0            & ~~~ &     S1                 & ~~~ &         G2              & ~~~ &         G3             \\
&&&&&&&&&& \\
           & ~ &   $kT$         & ~ & $5.0^{+0.4}_{-0.3}$    & ~~~ & $5.4^{+0.5}_{-0.5}$    & ~~~ &  $4.5^{+0.8}_{-0.6}$    & ~~~ & $4.5^{+0.6}_{-0.5}$    \\ 
           & ~ &   $Z$          & ~ & $0.29^{+0.10}_{-0.10}$ & ~~~ & $0.14^{+0.11}_{-0.11}$ & ~~~ &  $0.20^{+0.23}_{-0.16}$ & ~~~ & $0.39^{+0.24}_{-0.20}$ \\ 
 MODEL A   & ~ &   $N_{\rm H}$  & ~ & $0.31$                 & ~~~ & $0.31$                 & ~~~ &  $0.31$                 & ~~~ & $0.31$               \\ 
           & ~ &   $\chi^2$/DOF & ~ & 178.0/170              & ~~~ & 138.1/138              & ~~~ &  94.0/83                & ~~~ & 123.2/104              \\ 
&&&&&&&&&& \\                                                                                                                                               

           & ~ &   $kT$         & ~ & $4.2^{+0.4}_{-0.3}$    & ~~~ & $4.5^{+0.6}_{-0.5}$    & ~~~ &  $4.0^{+1.0}_{-0.8}$    & ~~~ & $4.2^{+0.9}_{-0.7}$    \\ 
           & ~ &   $Z$          & ~ & $0.30^{+0.10}_{-0.09}$ & ~~~ & $0.16^{+0.10}_{-0.10}$ & ~~~ &  $0.24^{+0.26}_{-0.21}$ & ~~~ & $0.41^{+0.25}_{-0.20}$ \\ 
 MODEL B   & ~ &   $N_{\rm H}$  & ~ & $0.89^{+0.23}_{-0.22}$ & ~~~ & $0.89^{+0.29}_{-0.28}$ & ~~~ &  $< 3.0$                & ~~~ & $< 2.2$               \\ 
           & ~ &   $\chi^2$/DOF & ~ & 157.7/169              & ~~~ & 125.1/137              & ~~~ &  92.5/82                & ~~~ & 122.4/103              \\ 
&&&&&&&&&& \\

           & ~ &   $kT$         & ~ & $5.3^{+2.3}_{-1.3}$     & ~~~ & $5.5^{+4.1}_{-1.4}$    & ~~~ &  $7.1^{+3.1}_{-3.8}$    & ~~~ &  $4.6^{+5.5}_{-1.0}$   \\ 
           & ~ &   $Z$          & ~ & $0.32^{+0.10}_{-0.10}$  & ~~~ & $0.16^{+0.11}_{-0.10}$ & ~~~ &  $0.34^{+0.25}_{-0.27}$ & ~~~ &  $0.43^{+0.35}_{-0.22}$  \\ 
 MODEL C   & ~ &   $N_{\rm H}$  & ~ & $1.63^{+0.25}_{-0.25}$  & ~~~ & $1.44^{+0.52}_{-0.75}$ & ~~~ &  $2.5^{+1.2}_{-2.2}$    & ~~~ &  $<3.3$                  \\ 
           & ~ &   ${\dot M}$   & ~ & $<3550$                 & ~~~ & $<3500$                & ~~~ &  $<2000$                & ~~~ &  $<2390$                 \\ 
           & ~ &   $\chi^2$/DOF & ~ & 155.6/168               & ~~~ & 123.8/136              & ~~~ &  91.6/81                & ~~~ &  122.3/102               \\ 
&&&&&&&&&& \\

           & ~ &   $kT$         & ~ & $5.2^{+2.3}_{-0.8}$    & ~~~ & $6.6^{+3.1}_{-1.9}$    & ~~~ & $6.9^{+3.2}_{-3.7}$     & ~~~ & $5.1^{+5.0}_{-1.6}$    \\ 
           & ~ &   $Z$          & ~ & $0.33^{+0.10}_{-0.10}$ & ~~~ & $0.17^{+0.11}_{-0.11}$ & ~~~ & $0.34^{+0.26}_{-0.26}$  & ~~~ & $0.46^{+0.33}_{-0.24}$   \\ 
 MODEL D   & ~ &   $N_{\rm H}$  & ~ & $0.31$                 & ~~~ & $0.31$                 & ~~~ & $0.31$                  & ~~~ & $0.31$                   \\ 
           & ~ &   ${\dot M}$   & ~ & $2290^{+1040}_{-960}$  & ~~~ & $2500^{+750}_{-1100}$  & ~~~ & $<2320$                 & ~~~ & $<2670$                  \\ 
           & ~ & $\Delta N_{\rm H}$  & ~ & $3.8^{+4.2}_{-1.5}$    & ~~~ & $2.6^{+4.0}_{-1.0}$    & ~~~ & U.C.                    & ~~~ & U.C.                     \\ 
           & ~ &   $\chi^2$/DOF & ~ & 154.1/168              & ~~~ & 124.0/136              & ~~~ & 91.4/81                 & ~~~ & 122.3/102                \\ 
\hline                                                                                                                                                         
&&&&&&&&&& \\                                                                                                                                               

\end{tabular}
\end{center}

\parbox {7in}
{Notes: The best-fit parameter values and 90 per cent ($\Delta \chi^2 = 2.71$)
confidence limits from the spectral analysis of the ASCA data for E1455+223.
Details as for Table 5.}
\end{table*}

\begin{figure}
\centerline{\hspace{2.3cm}\psfig{figure=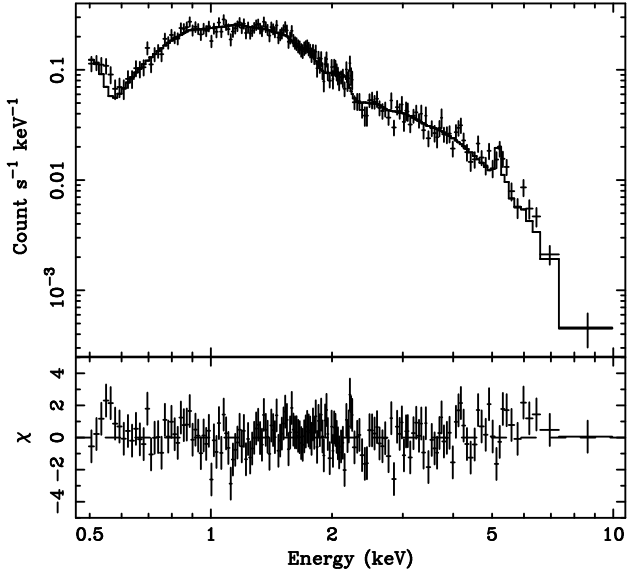,width=0.75\textwidth,angle=270}}
\caption{ (Upper Panel) The S0 spectrum for Zwicky 3146 with the best-fit 
solution for Model B overlaid. The data have been binned-up by a factor 5
along the energy axis for display purposes. 
(Lower Panel) Residuals to the fit. The positive residuals at $E \sim 0.55$ keV are 
due to small systematic uncertainties in response matrix around the oxygen edge in the detector.
}
\end{figure}

\begin{figure}
\centerline{\hspace{2.3cm}\psfig{figure=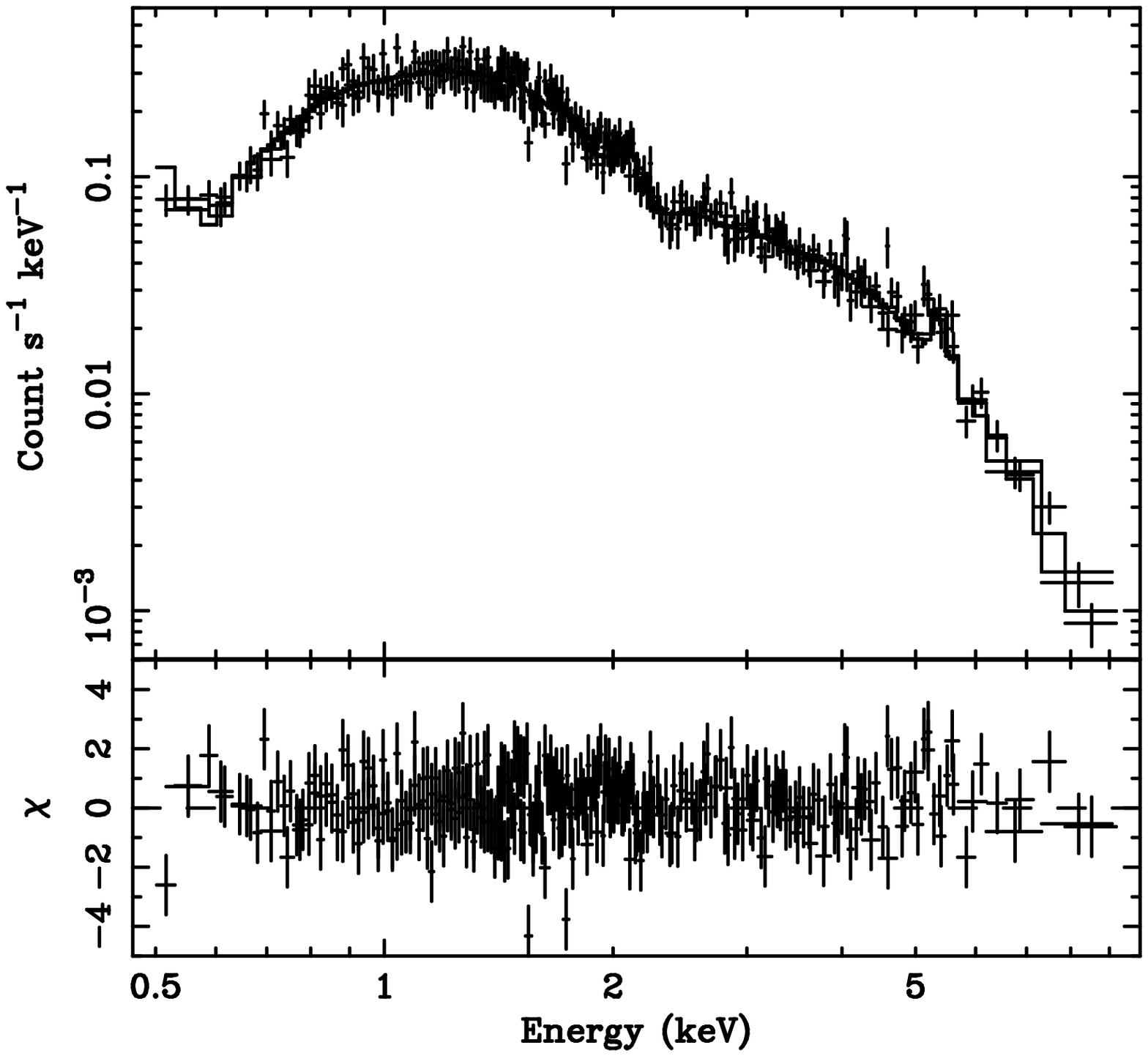,width=0.75\textwidth,angle=270}}
\caption{ (Upper Panel) The S0 spectrum for Abell 1835 with the best-fit 
solution for Model B overlaid. (Lower Panel) Residuals to the fit. Details as in Fig. 4.}
\end{figure}

\begin{figure}
\centerline{\hspace{2.3cm}\psfig{figure=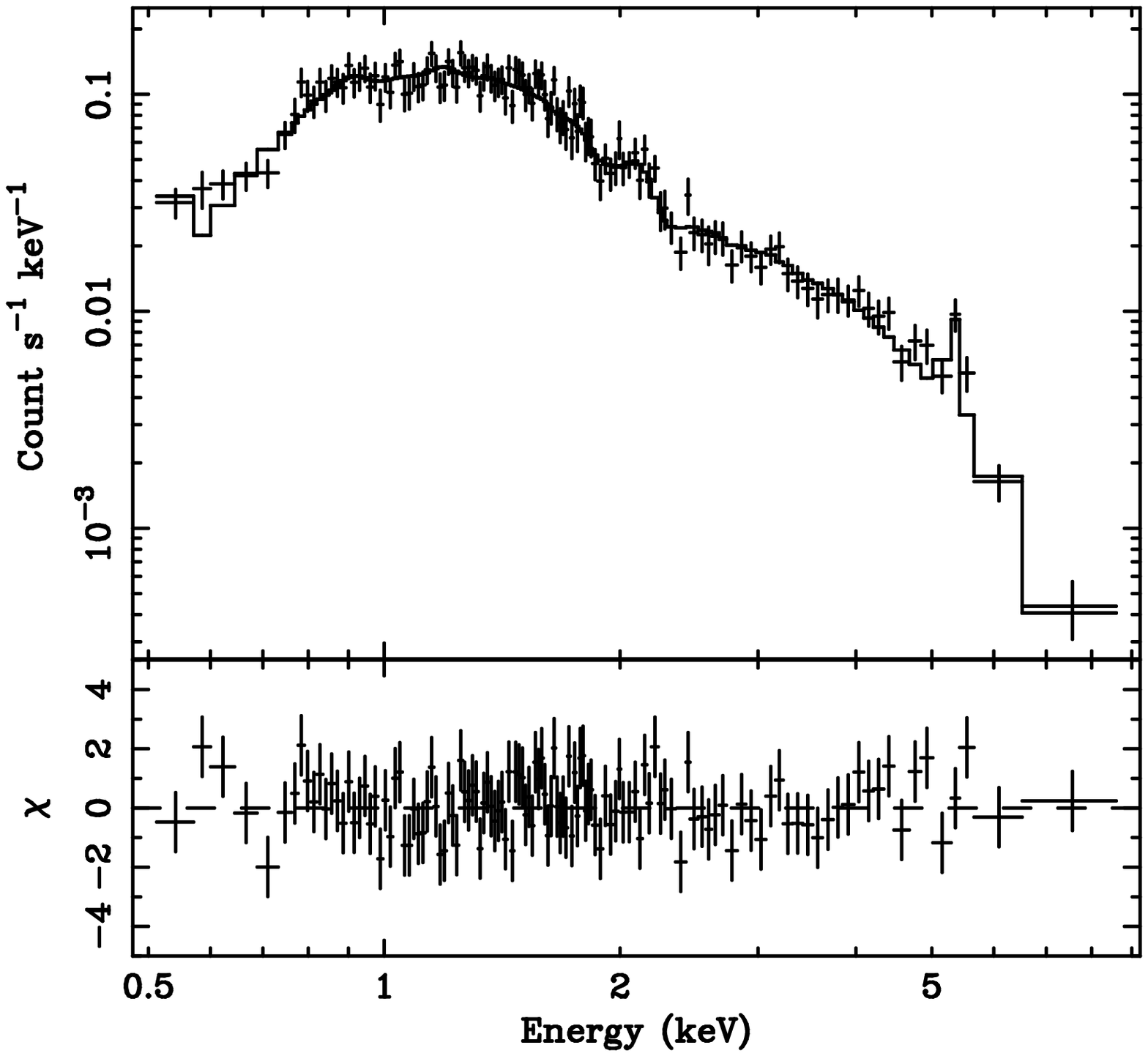,width=0.75\textwidth,angle=270}}
\caption{ (Upper Panel) The S0 spectrum for E1455+223  with the best-fit 
solution for Model B overlaid. (Lower Panel) Residuals to the fit. Details as in Fig. 4.}
\end{figure}

The ASCA spectra were first examined with a simple single-phase  model
consisting of an RS component, to account for the  X-ray emission from the
cluster,  and a  photoelectric absorption component (Morrison \& McCammon 1983)
normalized to the equivalent Galactic hydrogen column density 
along the line-of-sight to the cluster. The
free parameters in this model (hereafter Model A) were the  temperature,
metallicity and emission measure of the X-ray gas. 
The redshift
of the X-ray emission from the  cluster was fixed at the optically-determined
values for the CCGs (Table 2). 

We then examined a second model (Model B) in which the absorbing
column density was also allowed to be a free parameter in the fits.  
The fits to Abell 1835 and E1455+223 in particular showed 
highly significant improvements with the introduction of this single
extra fit parameter.  
(Note that absorbing material was assumed to lie at zero redshift in this model).
The best-fit parameter values and 90 per cent
($\Delta \chi^2 = 2.71$) confidence limits from the analyses 
with the single-phase models (A and B) are summarized in Tables 5--8. 

Although the single-phase modelling can provide  a useful parameterization of
the properties of the cluster gas, the results obtained with such a model 
should be interpreted with caution.  The deprojection analyses presented 
in Section 5
show that all three of the clusters discussed in this paper contain large cooling
flows.  The gas in the central regions of these clusters must therefore be
highly multiphase \ie contain a wide range of densities and temperatures at all 
radii. 

We therefore next examined the data with more sophisticated spectral  models in which 
the spectrum of the cooling flow was accounted for explicitly. The first of
these  models
(Model C) consists of an RS component  (to model the emission from  the ambient
ICM in the region of interest) and a cooling-flow component (following the
models of Johnstone \etal 1992) modelling the X-ray spectrum of gas cooling
from the ambient cluster temperature, to temperatures below the X-ray waveband, 
at constant pressure. Note that Model C introduces only one
extra free parameter into the fits relative to the single-phase Model B; the
mass deposition rate of cooling gas.  The upper temperature of the cooling gas,
the metallicity, and the absorbing column density acting on the cooling flow were 
tied to those of the ambient cluster emission modelled by the RS component. 

Fourthly, we examined a further cooling-flow model (D; which we expect to be 
the most physically-appropriate model) in which an
intrinsic X-ray absorbing column density was associated with the cooling-flow. 
The excess absorption is modelled as a uniform absorbing screen in front of 
the cooling flow,  at the redshift of the cluster, with the column
density a free parameter in the fits.  The
column density acting on the ambient cluster emission was fixed at the Galactic 
value.  The best-fit parameter values and confidence limits obtained with the 
multiphase, cooling-flow models (C,D) are also summarized in Tables 5--8.

Finally, a fifth model in which the  X-ray emission from the cluster was 
parameterized by a combination of two  RS components was studied. However, the
statistical significance of including the second RS component in the fits (2
extra free parameters) is low and the results on  the temperatures and emission
measures  of the two  components, which  are poorly  constrained by the
data, are not presented here.

\subsection{Results from the spectral analysis}

\begin{table*}
\vskip 0.2truein
\begin{center}
\caption{All instruments combined together }
\vskip 0.2truein
\begin{tabular}{ c c c c c c c c c c c  }
&&&&&&&&&&  \\                                                                                                                                               
\hline                                                                                                                                                         

            & ~ &   Parameters   & ~ &     Model A            & ~~~ &     Model B            & ~~~ &     Model C            & ~~~ &    Model D              \\
&&&&&&&&&&  \\                                                                                                                                               
            & ~ &   $kT$         & ~ & $5.74^{+0.19}_{-0.18}$ & ~~~ & $6.07^{+0.33}_{-0.33}$ & ~~~ & $6.9^{+2.7}_{-0.9}$    & ~~~ &  $6.6^{+1.1}_{-0.7}$    \\ 
            & ~ &   $Z$          & ~ & $0.26^{+0.05}_{-0.04}$ & ~~~ & $0.26^{+0.04}_{-0.05}$ & ~~~ & $0.27^{+.05}_{-0.05}$  & ~~~ &  $0.27^{+0.05}_{-0.05}$ \\ 
Zwicky 3146 & ~ &   $N_{\rm H}$  & ~ & $0.34$                 & ~~~ & $0.17^{+0.12}_{-0.09}$ & ~~~ & $0.54^{+0.34}_{-0.29}$ & ~~~ &  $<1.02$                \\ 
            & ~ &   ${\dot M}$   & ~ &  ---                   & ~~~ &       ---              & ~~~ & $1870^{+1270}_{-1350}$ & ~~~ & $1330^{+1220}_{-820}$                  \\ 
            & ~ &   $\chi^2$/DOF & ~ & 864.4/871              & ~~~ & 858.6/870              & ~~~ & 852.3/866              & ~~~ &   853.6/866              \\ 
&&&&&&&&&&  \\                                                                                                                                                                            
            & ~ &   $kT$         & ~ & $8.41^{+0.38}_{-0.39}$ & ~~~ & $7.03^{+0.34}_{-0.33}$ & ~~~ & $9.1^{+5.3}_{-1.6}$    & ~~~ &  $9.5^{+1.3}_{-1.7}$    \\ 
            & ~ &   $Z$          & ~ & $0.26^{+0.05}_{-0.05}$ & ~~~ & $0.26^{+0.06}_{-0.05}$ & ~~~ & $0.30^{+0.06}_{-0.06}$ & ~~~ &  $0.31^{+0.06}_{-0.05}$ \\ 
Abell 1835  & ~ &   $N_{\rm H}$  & ~ & $0.22$                 & ~~~ & $0.72^{+0.11}_{-0.10}$ & ~~~ & $1.12^{+0.24}_{-0.34}$ & ~~~ &  $3.8^{+1.6}_{-0.4}$    \\
            & ~ &   ${\dot M}$   & ~ &  ---                   & ~~~ &        ---             & ~~~ & $2050^{+1280}_{-1440}$ & ~~~ & $2090^{+630}_{-700}$                  \\ 
            & ~ &   $\chi^2$/DOF & ~ & 976.4/958              & ~~~ & 909.2/957              & ~~~ & 898.7/953              & ~~~ &  891.8/953              \\ 
&&&&&&&&&&  \\                                                                                                                                                                            
            & ~ &   $kT$         & ~ & $5.01^{+0.26}_{-0.26}$ & ~~~ & $4.29^{+0.25}_{-0.24}$ & ~~~ & $5.0^{+2.6}_{-0.7}$    & ~~~ &  $5.4^{+1.9}_{-0.7}$    \\ 
            & ~ &   $Z$          & ~ & $0.23^{+0.06}_{-0.07}$ & ~~~ & $0.25^{+0.07}_{-0.06}$ & ~~~ & $0.26^{+0.07}_{-0.06}$ & ~~~ &  $0.27^{+0.07}_{-0.07}$ \\ 
E1455+223   & ~ &   $N_{\rm H}$  & ~ & $0.31$                 & ~~~ & $0.90^{+0.17}_{-0.16}$ & ~~~ & $1.48^{+0.45}_{-0.49}$ & ~~~ &  $3.8^{+2.0}_{-0.8}$    \\ 
            & ~ &   ${\dot M}$   & ~ &  ---                   & ~~~ &        ---             & ~~~ & $1890^{+1300}_{-1490}$  & ~~~ & $2030^{+720}_{-880}$                  \\ 
            & ~ &   $\chi^2$/DOF & ~ & 541.8/501              & ~~~ & 503.5/500              & ~~~ & 498.8/496              & ~~~ &  498.2/496              \\ 
\hline                                                                                                                                                   
&&&&&&&&&&  \\

\end{tabular}
\end{center}
 
\parbox {7in}
{ Notes: The best-fit parameter values and 90 per cent  ($\Delta \chi^2 =
2.71$) confidence limits from the spectral analyses with data from all 4 detectors 
combined. Temperatures ($kT$) are in keV and metallicities ($Z$) are quoted  as a
fraction of the solar Value (Anders \& Grevesse 1989). Column densities
($N_{\rm H}$) are in units of $10^{21}$ atom cm$^{-2}$. $kT$, $Z$, 
and $N_{\rm H}$ are linked in the fits. However, the mass deposition
rates for each detector (\Mdot; quoted in \Msunpyr) are included as independent fit
parameters, due to variations in source extraction area and uncertainties in the flux 
calibration of the instruments. 
The \Mdot~value quoted for models C and D is for the S0 detector. 

}
\end{table*}

The results from the spectral analysis,  presented in Tables 5--8,   provide a
consistent description of the X-ray properties of the clusters.  We find
good agreement in the results from the different detectors. Interestingly, 
in a reduced $\chi^2$ sense, all four
spectral models provide a statistically  adequate description of the  ASCA spectra. 
Even with the single-phase models, however, the statistical 
improvement obtained by allowing the X-ray absorption to fit freely (\ie the 
improvement obtained with Model B over Model A) is very high -- particularly for 
Abell 1835 and E1455+223 where a simple $F$-test (Bevington
1969) indicates the improvement to be significant at $ >> 99.9$ per cent 
confidence.  Model B indicates excess column densities 
(assumed to lie at zero redshift) 
in Abell 1835 and E1455+223 of $5.0^{+1.1}_{-1.1}
\times 10^{20}$ \apc and $5.9^{+1.7}_{-1.6} \times 10^{20}$ \apc, respectively. The data for 
Zwicky 3146 prefer a column density marginally less than the nominal Galactic value 
(Stark \etal 1992). 

The introduction of the cooling flow component into the fits with Model C 
results in a further reduction in $\chi^2$. However, the statistical
significance of this improvement, relative to the model B results, is marginal, being required 
at $> 95$ per cent confidence 
only with the Abell 1835 data. (Note, however, that the improvement obtained with model C with respect to
model A is very high.) 
In general, the lowest $\chi^2$ values are obtained with Model D. 
However, it is difficult to interpret the improvement in $\chi^2$ obtained 
with Model D with respect to Models A--C 
in terms of a statistical significance since Model D
includes fit parameters and constraints not present in the other models. 
Within the context of Model D  the cooling flow component is required at high significance (Table 8). The data for 
Abell 1835 and E1455+223 also require significant amounts of intrinsic absorption associated with their cooling flows 
($N_{\rm H} = 3.8^{+1.6}_{-0.4} \times 10^{21}$ \apc for Abell 1835 and $3.8^{+2.0}_{-0.8} \times 10^{21}$
\apc for E1455+223) whereas the data for Zwicky 3146 are consistent with Galactic absorption. 
Note that only the SIS data have the spectral resolution and sensitivity at lower
energies ($E \approxlt 1$ keV) to detect the presence of cooling flows in the clusters. The GIS 
data do not provide firm constraints on the emission from cooling gas. 

We conclude that the ASCA spectra alone are unable to
discriminate  at high significance between the multiphase cooling flow 
models (C, D) and the
single-phase model (B) for the clusters. Although adopting  Model D  as intuitively the most
reasonable description of the X-ray emission from the clusters leads to a strong spectral requirement 
for large cooling flows in all three systems, it is only through the 
combination of the spectral results with the results from the deprojection analyses 
discussed in Section 5, that the presence of massive cooling flows in these clusters is 
firmly established.

Finally in this Section, we note the possible effects of uncertainties in
the low-energy calibration of the SIS data on our results. Our own analyses 
of ASCA observations of bright, nearby X-ray sources indicate that the current
GSFC response matrices (released 1994 November 9) may slightly
overestimate the low-energy response of the SIS instruments. This
systematic effect can lead to overestimates of Galactic column densities
by $1-3 \times 10^{20}$ \apc (see also the discussion of calibration
uncertainties associated with the ASCA instruments on the ASCA Guest
Observer Facility World Wide Web pages at
${http://heasarc.gsfc.nasa.gov/docs/asca/cal\_probs.html}$). Fixing the
Galactic column densities in our analyses with spectral Model D at values
$1.7 \times 10^{20}$ \apc in excess of the nominal Galactic values for the
clusters [$1.7 \times 10^{20}$ \apc being the best-fit systematic excess
column density determined from our analysis of ASCA observations of the
Coma cluster, which we expect to contain little or no intrinsic absorbing
material (White \etal 1991, Allen \& Fabian 1996), we determine best-fit 
intrinsic column densities for Zwicky 3146, Abell 1835 and E1455+223 of 
$N_{\rm H} < 0.6 \times 10^{21}$ \apc, $N_{\rm H} = 3.2^{+1.4}_{-0.6} 
\times 10^{21}$ \apc, and $N_{\rm H} = 3.1^{+1.9}_{-0.6} \times 10^{21}$ \apc, 
respectively. Hence, our conclusions on the presence of 
intrinsic absorbing material in these clusters are 
essentially unaffected by uncertainties in the calibration of the
SIS instruments. Note that the temperature constraints are also 
little-affected by the calibration uncertainties, with the best-fit
temperatures and 90 per cent confidence limits from the column 
density-adjusted fits being  $kT = 6.9^{+1.2}_{-0.7}$, $9.1^{+2.1}_{-1.3}$ and 
$5.2^{+2.2}_{-0.7}$ keV, respectively.

\subsection{Single-phase and Multiphase temperature results}

The most notable differences between the results obtained with the single-phase (Model B)
and multiphase (Models C,D) models are in the temperature determinations for the clusters. 
The cooling flow models imply significantly higher ambient cluster temperatures, and therefore 
larger integrated cluster masses.

The presence of a cooling flow will naturally lead to differences
between the mean emission-weighted and mass-weighted temperatures for a 
cluster.  The X-ray emissivity of the cooler, denser material in the cooling flow 
will be significantly higher than that of the surrounding hotter gas.  The mean
emission-weighted temperature will therefore be biased to
temperatures below the mass-weighted value for the system. The 
effects of emission-weighting become particularly important for 
observations made with the comparatively low spectral resolution and limited ($0.1-2.4$ keV)
bandpass of the ROSAT PSPC. In Table 9  we present
the results from a spectral analysis of the PSPC data for the central 6 arcmin
radius regions of Zwicky 3146 and Abell 1835.  The single-phase models again provide a 
statistically adequate description of the spectra ($\chi^2_\nu \sim 1.0$)  but the measured emission weighted temperatures 
are only $3.2^{+1.4}_{-0.7}$ keV and $3.8^{+1.6}_{-0.9}$  keV respectively, 
much less than $6.6^{+1.1}_{-0.7}$ keV and $9.5^{+1.3}_{-1.7}$ keV determined from the multiphase analysis of the ASCA spectra 
(Model D). The importance 
of distinguishing between single-phase and multiphase models in the analysis of X-ray data for clusters 
is discussed in more detail in Section 7.1.

\begin{table}
\vskip 0.2truein
\begin{center}
\caption{PSPC spectra for Zwicky 3146 and Abell 1835}
\vskip 0.2truein
\begin{tabular}{ c c c c c }
&& \\
\hline                                                                                                                                                   

Parameter    & ~ &     Zwicky 3146          & ~ &     Abell 1835            \\
&& \\                                                                      
$kT$         & ~ &  $3.2^{+1.4}_{-0.7}$     & ~ &  $3.8^{+1.6}_{-0.9}$      \\ 
$Z$          & ~ &  $0.52^{+1.68}_{-0.48}$  & ~ &  $0.0^{+0.22}_{-0.0}$   \\ 
$N_{\rm H}$  & ~ &  $0.23^{+0.05}_{-0.06}$  & ~ &  $0.18^{+0.03}_{-0.02}$   \\ 
$\chi^2$/DOF & ~ &    26.3/22               & ~ &    17.0/20                \\ 
\hline                                                                                                                                                   
&& \\
\end{tabular}
\end{center}

\parbox {3.3in}
{ Notes: The best-fit parameter values and 90 per cent ($\Delta \chi^2 = 2.71$)
confidence limits from the spectral analysis  of the PSPC data. }
\end{table}

\section{Deprojection Analysis}

\subsection{General results} 

\begin{table}
\vskip 0.2truein
\begin{center}
\caption{Deprojection analyses of the clusters}
\vskip 0.2truein
\begin{tabular}{ c c c c c c c }
&&&&&& \\                                                  
\hline                                                                                                                                                   
            & ~ & $t_{\rm cool}$         & ~ & $r_{\rm cool}$    & ~ & ${\dot M}$    \\  
&&&&&& \\                                                                                                            
Zwicky 3146 & ~ & $1.01^{+0.06}_{-0.06}$ & ~ & $231^{+50}_{-37}$ & ~ & $1355^{+408}_{-129}$ \\
Abell 1835  & ~ & $1.36^{+0.31}_{-0.31}$ & ~ & $231^{+26}_{-13}$ & ~ & $1106^{+455}_{-425}$ \\
E1455+223   & ~ & $1.15^{+0.17}_{-0.15}$ & ~ & $213^{+47}_{-33}$ & ~ & $732^{+162}_{-64}$  \\
\hline                                                                                                                   
&&&&&& \\                                                  

\end{tabular}
\end{center}

\parbox {3.3in}
{ Notes: A summary of the results from the deprojection analyses of the  ROSAT
HRI data. Cooling times ($t_{\rm cool}$) are mean values for the central (8
arcsec) bin and are in units of $10^9$ \yr. Cooling radii ($r_{\rm cool}$), the radii at
which the cooling time exceeds the Hubble time ($1.3 \times 10^{10}$ yr),
are in \kpc. Integrated mass deposition rates within the cooling radii 
(${\dot M}$ ) are in units of \Msunpyr. Errors on the cooling times are the 10 and 90 percentile values from
100 Monte Carlo simulations. The upper and lower confidence limits on the
cooling radii are the points where the 10 and 90 percentiles exceed, and become
less than, the Hubble time, respectively. Errors on the mass deposition rates
are the 90 and 10 percentile values at the upper and lower limits for the
cooling radius.  Galactic column densities as listed in Table 1 are 
assumed. 
}
\end{table}

\begin{figure*}
\centerline{\hspace{3.2cm}\psfig{figure=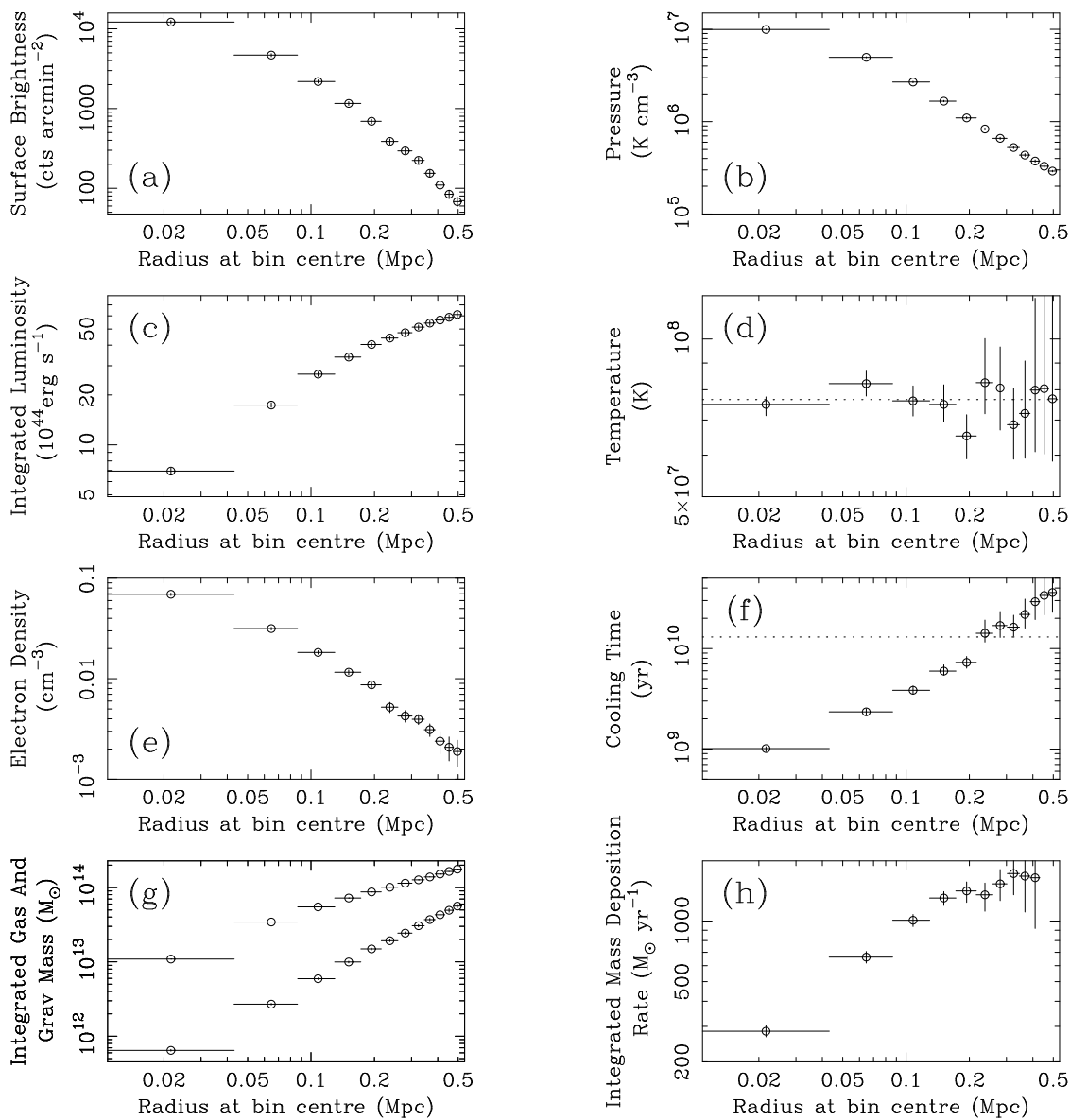,width=1.35\textwidth,angle=270}}
\caption{ A summary of the results from the deprojection analysis   of
the HRI data for Zwicky 3146. From left to right, top to bottom, we plot; (a)
surface brightness, (b) pressure, (c) integrated luminosity, (d) temperature,
(e) electron density, (f) cooling time, (g) integrated gas and gravitational
mass and  (h) integrated mass deposition rate. Data points are  mean values and
1$\sigma$ errors (in each radial bin) from 100 Monte Carlo simulations, except
for (d), (f) and (h) where the  median and 10 and 90 percentile values have
been  plotted.}
\end{figure*}

\begin{figure*}
\centerline{\hspace{3.2cm}\psfig{figure=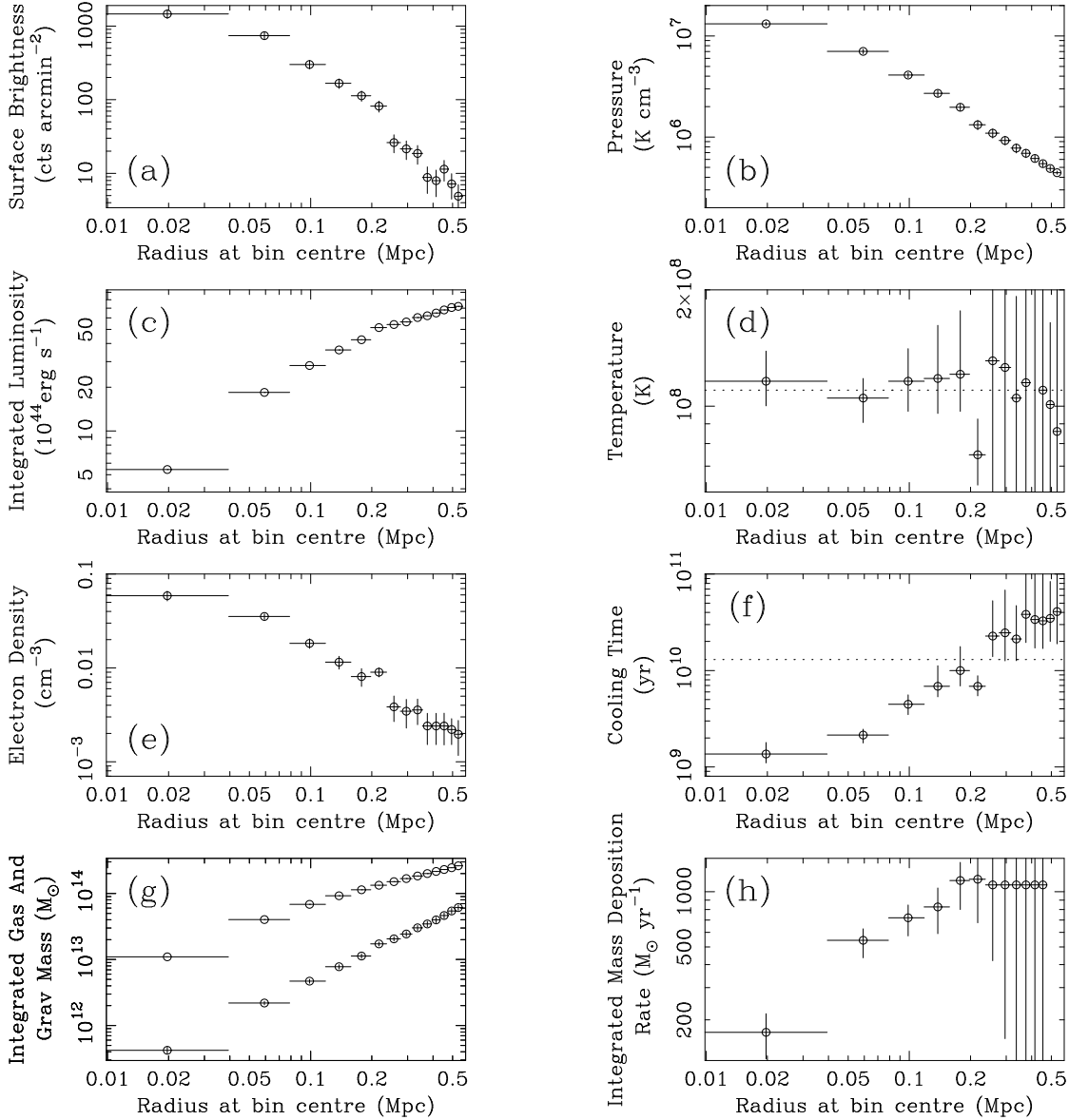,width=1.35\textwidth,angle=270}}
\caption{ A summary of the results from the deprojection analysis   of
the HRI data for Abell 1835.  Details as for Fig. 7}
\end{figure*}

\begin{figure*}
\centerline{\hspace{3.2cm}\psfig{figure=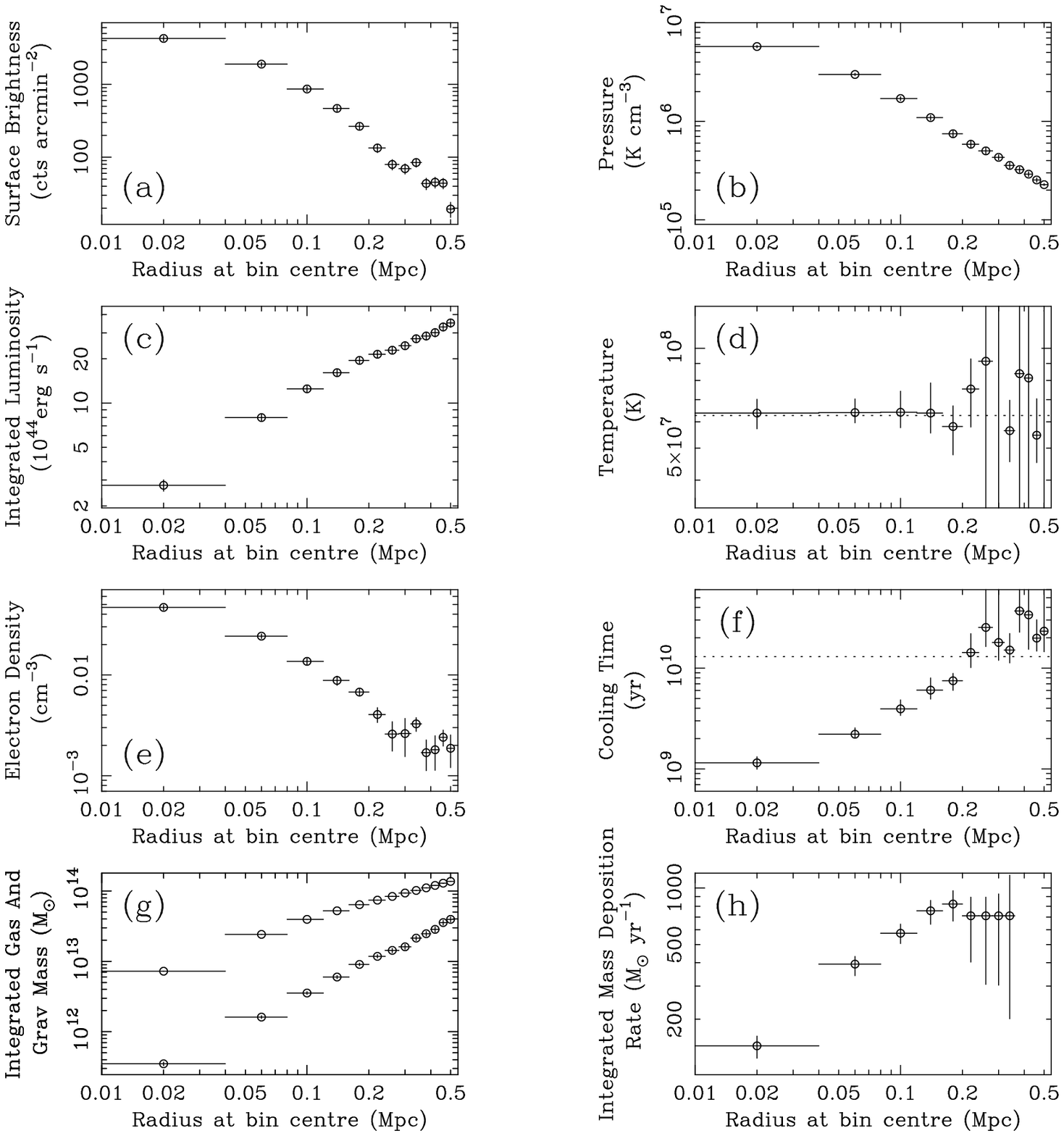,width=1.35\textwidth,angle=270}}
\caption{ A summary of the results from the deprojection analysis   of
the HRI data for E1455+225.  Details as for Fig. 7}
\end{figure*}

We have carried out a deprojection analysis of the ROSAT images 
using an updated version of the code of Fabian \etal (1981).  Using assumptions of spherical
symmetry and hydrostatic equilibrium in the ICM, the deprojection technique can
be used to study the properties of the intracluster gas  (\eg density,
pressure, temperature, cooling rate) as a function of radius. The deprojection
method requires that either the total mass profile (which defines the pressure
profile)  or the gas temperature profile be specified.   
Following ASCA observations of nearby cooling flow clusters (Fabian
\etal 1996), and the results from the combined X-ray and
gravitational lensing study of the cooling-flow cluster PKS0745-191 (Allen \etal 
1996), we assume that the mass-weighted  temperature profiles
in the clusters remain constant, at the temperatures determined from the fits with 
the multiphase spectral models to the combined detector data sets (Table 8).
Spectral model D is intuitively the preferred model, but using 
the full temperature range allowed by models C and D probably provides a more 
realistic estimate of the true uncertainty on the mass-weighted temperatures
in the highly complex, multiphase environments of the cooling flows.
Column densities were are fixed at the Galactic values from Stark \etal (1992),   
but see also Section 5.3.

The clusters discussed in this paper are remarkably similar in their X-ray 
properties to PKS0745-191 and the analogy to that system is a reasonable one. 
It should be noted that although the 
deprojection method of Fabian \etal (1981) is essentially a single-phase
technique, it produces results in good agreement with the more detailed
multi-phase treatment of Thomas, Fabian \& Nulsen (1987) and, due to its 
simple applicability at large radii, is better-suited to the present project.

The azimuthally averaged X-ray surface brightness profiles of the clusters
determined from the HRI data (background-subtracted and corrected for telescope
vignetting) and the results from the deprojection analyses are  summarized in 
Figs. $7-9$.   The primary  results on the cooling flows in the clusters; the
central cooling times, the cooling radii and the integrated mass deposition
rates within the cooling radii are listed in Table 10.  The results on the
cooling flow in Zwicky 3146 are in good  agreement with those  reported by Edge
\etal (1994) from an earlier analysis of the HRI data. The mass deposition from the 
cooling flows is distributed throughout the inner  $\sim 200$ kpc of the clusters with ${\dot M}
\approxpropto r$.  As noted in Section 1, such distributed mass deposition
profiles requires that the central ICM is inhomogeneous (Nulsen
1986; Thomas, Fabian \& Nulsen 1987;  Fabian 1994). Note also that the mass
deposition profiles shown in Figs. $7-9$(h)  flatten at radii $< 200$ kpc, 
where the cooling time is $\approxlt 10^{10}$ yr.  Thus accounting for the
look-back time to the clusters ($\sim 4 \times 10^{9}$ yr) does not
significantly alter the integrated mass deposition rates.

\subsection{Parameterization of the cluster masses} 


\begin{figure}
\centerline{\hspace{3.2cm}\psfig{figure=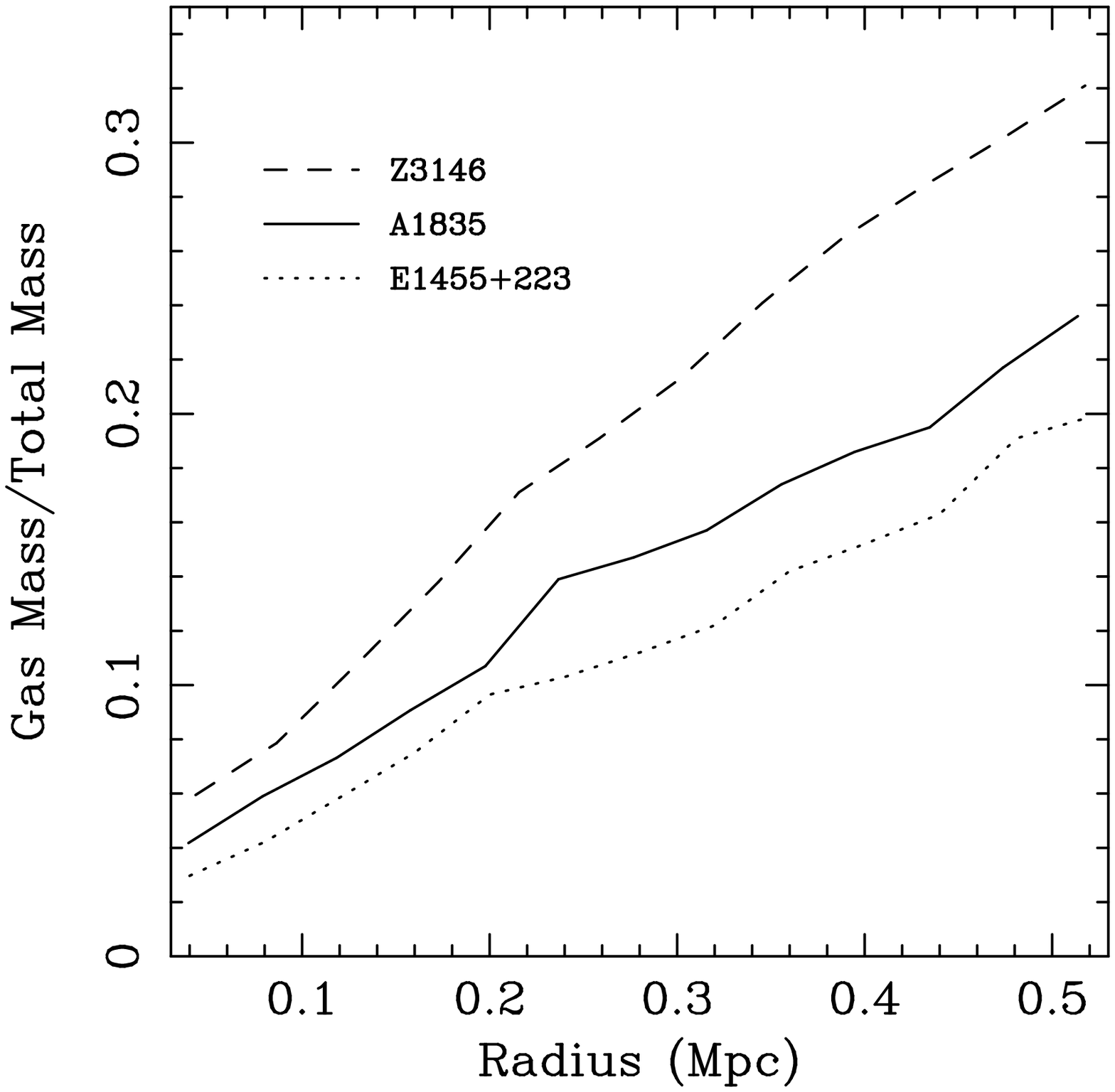,width=0.7\textwidth,angle=270}}
\caption{The ratio of the gas mass to the total mass as a function of radius.}
\end{figure}

\begin{table}
\vskip 0.2truein
\begin{center}
\caption{Cluster mass distributions}
\vskip 0.2truein
\begin{tabular}{ c c r l r l c }
\hline                                                                                                                   
\multicolumn{1}{c}{} &
\multicolumn{1}{c}{} &
\multicolumn{2}{c}{$kT$} &
\multicolumn{2}{c}{$\sigma$} &
\multicolumn{1}{c}{$r_c$} \\
&&&& \\                                                                            
Zwicky 3146 & ~ & $6.6^{+3.0}_{-0.7}$ & $(^{+1.1}_{-0.7})$  & $850^{+175}_{-50}$   & $(^{+60}_{-50})$  & 45     \\
Abell 1835  & ~ & $9.5^{+4.9}_{-2.0}$ & $(^{+1.3}_{-1.7})$  & $1000^{+300}_{-100}$ & $(^{+150}_{-70})$ & 50     \\
E1455+223   & ~ & $5.4^{+2.2}_{-1.1}$ & $(^{+1.9}_{-0.7})$  & $720^{+180}_{-70}$   & $(^{+150}_{-40})$ & 45     \\
\hline                                                                                                                   
&&&& \\                                                  

\end{tabular}
\end{center}

\parbox {3.3in}
{ Notes: A summary of the temperature constraints (in keV) and the 
velocity dispersions (in \kmps) and core radii (in kpc) of the
isothermal mass distributions (Binney \& Tremaine 1989) required to produce the 
flat temperature profiles shown in Figs. $7-9$(d). Errors on the temperatures give the 
the full range allowed by spectral models C and D. (The tighter 
constraints determined with model D alone are given in parentheses.) 
Errors on the velocity dispersions show the range of values required 
to match the temperature results.}
\end{table}

In Table 11 we summarize the mass distributions required to produce 
the flat temperature profiles
described in Section 5.1 We have parameterized 
the mass distributions as isothermal spheres
(Equation 4-125 of Binney \& Tremaine 1987) with core radii, $r_c$, and
velocity dispersions, $\sigma$. In Fig. 11 we
plot the X-ray gas mass/total mass ratios for the central 500 kpc of the
clusters (where the total mass is assumed to be described by the 
best-fit parameters listed in Table 11).

\subsection{The correction for intrinsic absorption} 

As discussed in Section 5.1, for the purposes of deprojection 
we have assumed that the column densities to the clusters are given by the 
Galactic values of Stark \etal (1992). However, the spectral analysis of Section 4 clearly 
shows that both Abell 1835 and E1455+223 exhibit significant excess absorption. 
This can have a significant effect on the mass deposition rates determined from the deprojection 
analyses. 

To correct the deprojection results for intrinsic absorption, we 
adopt spectral model D as the most reasonable description for the clusters. 
The intrinsic column densities in Abell 1835 and E1455+223 ($N_{\rm H} = 3.8 \times 10^{21}$
\apc; Table 8) are assumed to be in uniform screens in front of the cooling flows. The effects of
absorption on the observed HRI count rates from the cooling flows have been calculated from 
XSPEC simulations, using the ROSAT HRI response matrix issued by GSFC. For
both Abell 1835 and E1455+223, the intrinsic absorption acts to reduce the count rates 
from the cooling flows by a factor of two (in detail, factors 2.07 and 2.04, respectively). The true mass deposition rates from 
the cooling flows can therefore be assumed to be a factor 2 larger than the values 
listed in Table 10. [For Zwicky 3146, the maximum allowed intrinsic column density of $10^{21}$
\apc implies a maximum correction factor to the mass deposition rate of 1.28.] 
The absorption-corrected mass deposition rates for the clusters are summarized in Table 12. 
The agreement between these values and the spectrally-determined mass
deposition rates (Model D) is excellent. 

\section{Optical properties of the clusters} 

\subsection{The galaxy populations} 

The galaxy populations in Abell 1835 and E1455+223 have been studied using  the
Palomar B and I images of the clusters.  The SExtractor software of Bertin
(1995) was used to identify  galaxy candidates from their total (Kron)
I  magnitudes and  $(B-I)$ colours (measured in 3.0 arcsec diameter apertures from
seeing-matched  images). For Abell 1835, 283 objects were selected with 
$15.0<I<21.0$ and  $2.5<(B-I)<3.2$.  The CCG (which is bluer than allowed by this range)
was also included, making 284 galaxy identifications in total.  For E1455+223, 
188 objects with $15.0<I<21.0$ and $1.8<(B-I)<2.7$ were identified.  
Absolute V magnitudes were
calculated from the observed I magnitude,  and applying the appropriate
K-corrections.

In Figs. 11 (a), (b) we plot the (projected) galaxy profiles and the 
ratio of the galaxy mass to the total mass
for Abell 1835 and E1455+223. A value for (M/L)$_V$ of 10 has been assumed. 
Note the large central peaks in the ratio profiles [Fig. 12(b)] due to the 
luminous central galaxies, and the near-flatness of the ratios at large radii 
($r \approxgt 0.5$ Mpc) indicating that the galaxies follow an approximately isothermal 
distribution. 

\begin{figure}
\centerline{\hspace{3.2cm}\psfig{figure=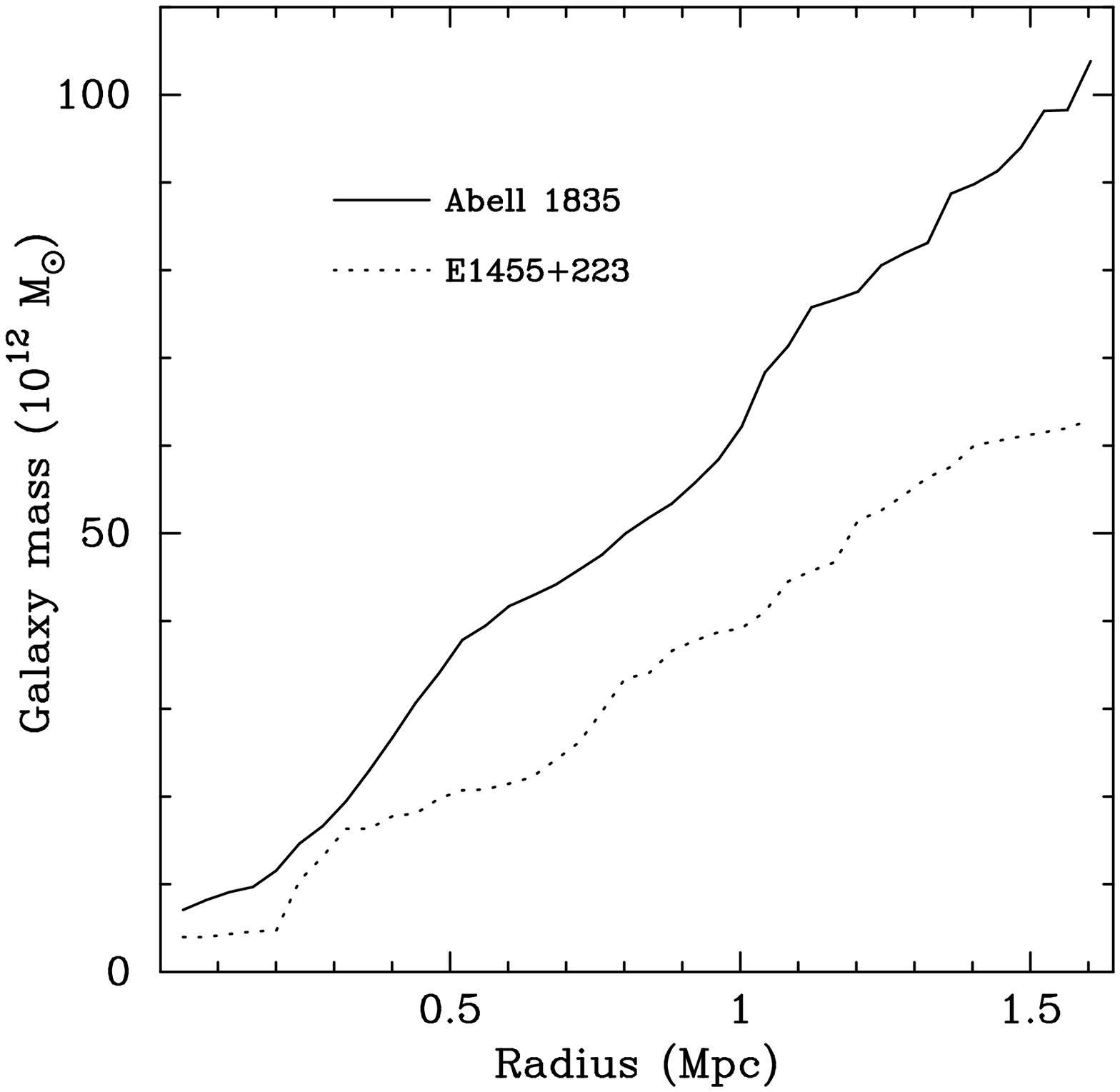,width=0.7\textwidth,angle=270}}
\vskip -0.2truein
\centerline{\hspace{3.2cm}\psfig{figure=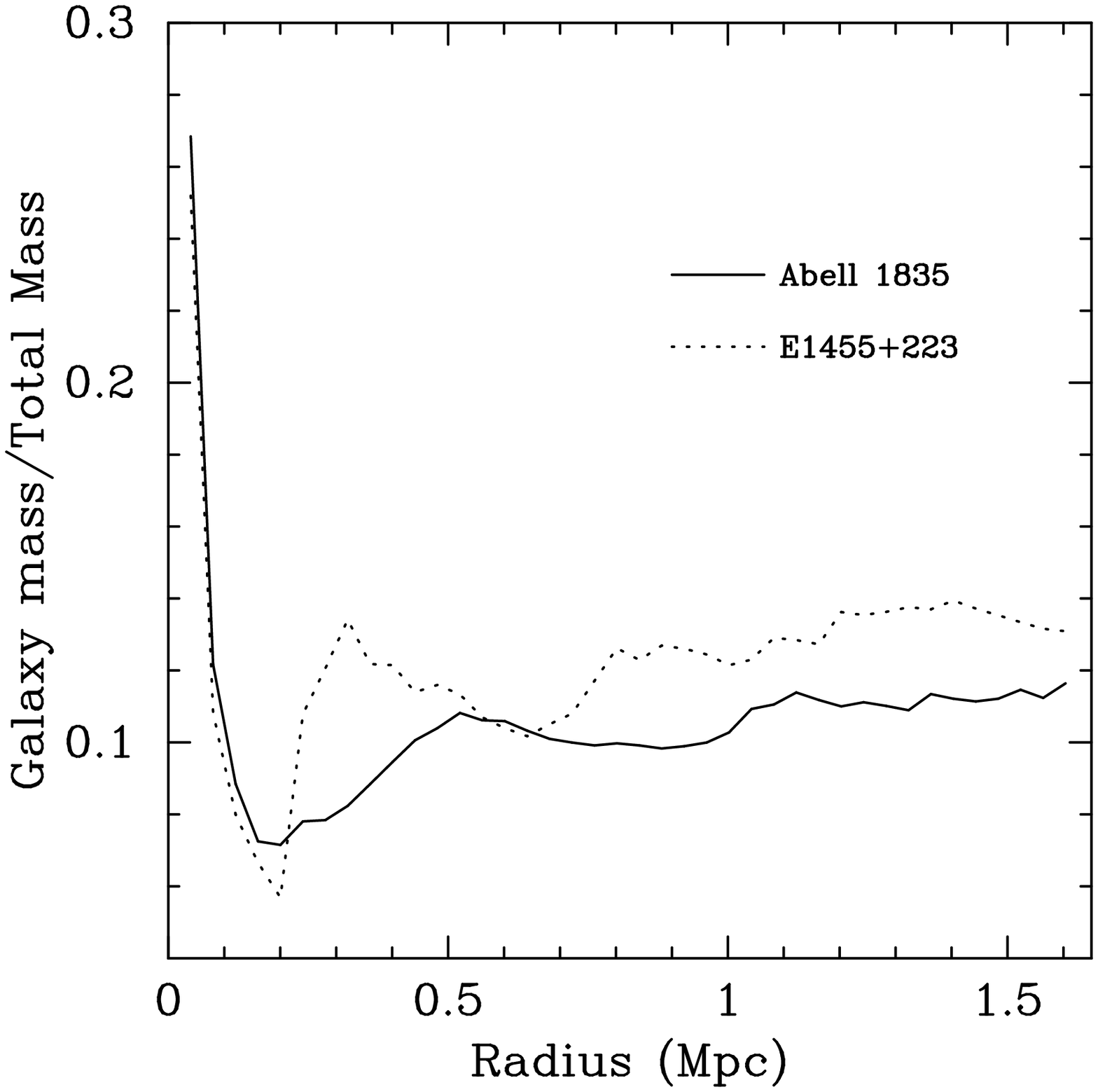,width=0.7\textwidth,angle=270}}
\caption{(a) The projected galaxy mass as a function of radius in  Abell 1835
and E1455+223. (b) The ratio of the galaxy mass to the total mass (derived using 
the temperature constraints from spectral Model D).} 
\end{figure}

\subsection{Spectra of the central cluster galaxies}

\begin{figure*}
\centerline{\hspace{0.0cm}\psfig{figure=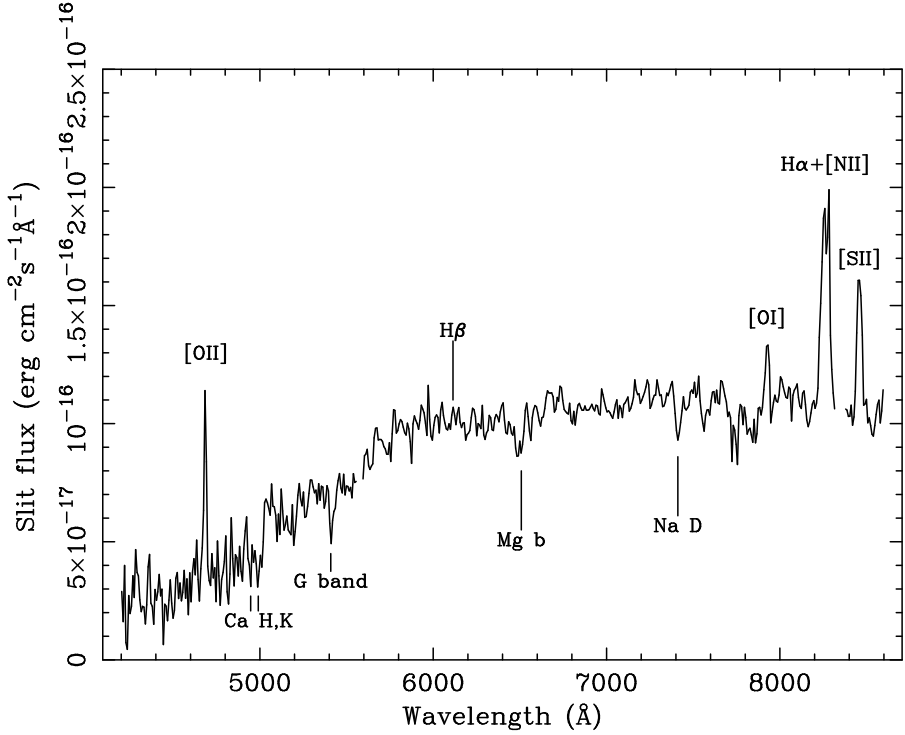,width=0.8\textwidth,angle=270}}
\caption{ Optical spectrum of the CCG of E1455+223 obtained with the Faint
Object Spectrograph on the INT in June 1991.}
\end{figure*}

Optical spectra for the CCGs of Zwicky 3146 and Abell 1835  are presented and
discussed by Allen (1995). These CCGs are two of the most 
optically line-luminous central galaxies known, with luminosities in 
H$\alpha\lambda$6563 emission alone of $\sim 10^{43}$ \ergps.  
In Fig. 13 we present the  spectrum of the CCG in
E1455+223. The data were obtained with the Faint Object Spectrograph on the
Isaac Newton Telescope (INT) in June 1991. A slit width of 1.5 arcsec was used,
providing a resolution of $\sim 16$ \AA~in the first spectral  order.

The CCG of E1455+223 exhibits strong, narrow emission 
lines (FWHM $\sim 500$ \kmps) and an enhanced blue continuum with respect to a typical  elliptical
galaxy spectrum. Both features are characteristic properties of CCGs in large
cooling flows (Johnstone, Fabian \& Nulsen
1987; Heckman \etal 1992; McNamara \& O'Connell 1989; Crawford \& Fabian 1993; Allen 1995). 
The observed flux in H$\alpha\lambda$6563 of $1.9 \pm 0.1 \times
10^{-15}$ \ergpcmsqps implies a slit luminosity of  $6.1 \pm 0.3 \times
10^{41}$ \ergps. However, the H$\beta\lambda$4861 emission line is only 
marginally detected and, after correcting for 
absorption by the underlying stellar continuum (a factor $\sim 2$ correction), 
we derive an H$\alpha\lambda$6563/H$\beta\lambda$4861
flux ratio of $\approxgt 4.0$. This implies significant intrinsic reddening 
at the source [$E(B-V) \approxgt 0.3$] and an intrinsic 
H$\alpha\lambda$6563 (slit) luminosity of $\approxgt 10^{42}$ \ergps. 
[The reddening laws of Seaton (1979) and Howarth (1983) have been
used.] Donahue, Stocke and Gioia (1992) present results from a narrow-band
H$\alpha$+[NII] imaging study of E1455+223. These authors show that the
line emission from the CCG is highly extended and that the total
line flux exceeds the slit flux reported here by a factor $\sim 7-8$.
The Donahue \etal (1992) results, together with the spectral results
reported here, thus imply an H$\alpha\lambda$6563 luminosity for
the CCG of E1455+223 of $\approxgt 7 \times 10^{42}$ \ergps,
comparable to that of Zwicky 3146 and Abell 1835.

\begin{table*}
\vskip 0.2truein
\begin{center}
\caption{Reddening and Excess absorption}
\vskip 0.2truein
\begin{tabular}{ c c c c c c c }                                                                      
\hline                                                                                                                   
              & ~ & $N_{{\rm H, X-ray}}$ & Spectral ${\dot M}$   & Corrected deproj. ${\dot M}$   & $E(B-V)$      & $N_{{\rm H, OPT}}$    \\
&&&&&& \\                                                                                                                                                        
Zwicky 3146   & ~ &       $<1.0$         & $1330^{+1220}_{-820}$ & $1355^{+637}_{-161}$  & $0.22^{+0.15}_{-0.14}$ & $1.3^{+0.9}_{-0.8}$   \\
Abell 1835    & ~ & $3.8^{+1.6}_{-0.4}$  & $2090^{+630}_{-700}$  & $2291^{+943}_{-881}$  & $0.49^{+0.17}_{-0.15}$ & $2.8^{+1.0}_{-0.9}$   \\
E1455+223     & ~ & $3.8^{+2.0}_{-0.8}$  & $2040^{+720}_{-880}$  & $1491^{+330}_{-130}$  & $\approxgt 0.3$        & $\approxgt 1.7 $   \\
\hline                                                                                                                                                              
&&&&&& \\                                                                                                                                                        

\end{tabular}
\end{center}

\parbox {7in}
{Notes: Columns 2 and 3 list the intrinsic X-ray column densities
(without account for systematic uncertainties in the SIS calibration;
Section 4.3) and mass deposition rates determined from the spectral
analysis of the combined instrument data sets (Table 8). Column 4 lists the 
mass deposition rates, determined by deprojection, corrected a posteriori for 
the effects of X-ray absorption (Section 5.3). For Zwicky 3146, zero intrinsic 
absorption has been assumed. Applying the maximum allowed correction factor 
for this cluster (1.28) gives a mass deposition rate of $1734^{+815}_{-206}$. 
$E(B-V)$ estimates in column 5 are determined from the 
H$\alpha$6563/H$\beta$4861  line ratios in the CCGs. The data for Zwicky 3146 
and Abell 1835 are from Allen (1995). $N_{\rm H, OPT}$ values are the column 
densities of X-ray absorbing material implied by the $E(B-V)$ estimates,
following  the relation of Bohlin, Savage \& Drake (1978). } 
\end{table*}

The CCGs of Zwicky 3146 and Abell 1835 also exhibit 
significant intrinsic reddening.
In Table 12 we summarize the results on intrinsic X-ray absorption and optical
reddening for the clusters. These results, together with the optical/X-ray/UV results
discussed by Allen \etal (1995) for a larger sample of cooling flows at 
intermediate redshifts ($z \sim 0.15$), 
indicate an interesting tendency for clusters with large column densities of 
intrinsic X-ray absorbing material to 
exhibit significant intrinsic reddening. The similarity of the column densities
inferred from the X-ray, optical and UV data, across a 
variety of aperture sizes, suggests a dust-to-gas ratio in these galaxies similar to that
in our own Galaxy. The results also suggest that much of the dust 
may be associated with, or entrained within, the X-ray absorbing gas.

\section{Discussion}

We have discussed, in detail, the X-ray properties of Zwicky 3146, Abell 1835
and E1455+223. We have shown that all three of these clusters contain 
exceptionally large cooling flows. 
Zwicky 3146 and Abell 1835 are amongst the most X-ray luminous clusters 
known (Table 2). The cooling flows in these systems account for $\sim 15-20$ per cent
of the total intrinsic luminosity in the $2-10$ keV band, and as much as $\sim 40$ per cent in the 
$0.1 - 2.4$ keV ROSAT band. 
With E1455+223, which is a factor $2-3$ less luminous than the other clusters, 
the cooling flow accounts for $\sim 35$ per cent of the luminosity in 2-10 keV band 
and $\sim 60$ per cent of the emission between 0.1 and 2.4 keV. 

Both Abell 1835 and E1455+223 exhibit significant intrinsic absorption in their ASCA spectra. 
The need for excess absorption is found 
in both the single-phase and multiphase (cooling flow) spectral analyses and cannot 
reasonably be attributed to uncertainties in the Galactic column densities (Stark \etal 1992).
The most plausible interpretation of the excess absorption is that it is due
to material associated with the cooling flows (spectral Model D).
The mass of absorbing gas implied by an intrinsic column density of $\sim 3.8 \times 10^{21}$ \apc, 
distributed in a uniform (circular) screen across the central 220 kpc (radius $\sim r_{\rm cool}$) of the 
clusters, is $\sim 4.6 \times 10^{12}$ \Msun. Such a mass could plausibly be accumulated by the 
cooling flows in Abell 1835 and E1455+223 in $\sim 2-3 \times 10^{9}$ yr. [See also White \etal (1991) and 
Allen \etal (1993).] Note also the excellent
agreement in the mass deposition rates for the cooling flows  
determined with the spectral and deprojection methods (Section 5.3) under this assumption for the 
distribution of absorbing gas. 
With spectral models B and C, wherein the excess absorption is assumed to cover 
the whole cluster, the mass of absorbing gas is implausibly high. 
The intrinsic column densities inferred for Abell 1835 and E1455+223 are similar to those observed in nearby 
cooling flows (White \etal 1991; Allen \etal 1993; Fabian \etal 1994; Fabian \etal 1996). 
Note also that the metallicities of $Z \sim 0.25 -0.30 Z_\odot$ measured for these clusters 
are similar to those observed in 
nearby systems, and imply that the the bulk of the enrichment of the ICM in
these clusters occurred before redshifts of $\sim 0.3$.

\subsection{Cooling flows and multiphase models}

\begin{table*}
\vskip 0.2truein
\begin{center}
\caption{Comparison of multiphase and single-phase results}
\vskip 0.2truein
\begin{tabular}{ c c c c c c c c }
\hline                                                                                                                                                              
\multicolumn{1}{c}{} &
\multicolumn{1}{c}{} &
\multicolumn{2}{c}{PSPC (SP)} &
\multicolumn{2}{c}{ASCA (SP)} &
\multicolumn{2}{c}{ASCA (MP)} \\
 Cluster    & ~ & $kT$   & $\chi^2_\nu$ ($\nu$)   & $kT$ & $\chi^2_\nu$ ($\nu$)          & $kT$   & $\chi^2_\nu$ ($\nu$) \\
&&&&&&& \\
Zwicky 3146 & ~ & $3.2^{+1.4}_{-0.7}$ & 1.20 (22) & $6.1^{+0.3}_{-0.3}$ & 0.987 (870) & $6.6^{+1.1}_{-0.7}$ & 0.986 (866) \\
Abell 1835  & ~ & $3.8^{+1.6}_{-0.9}$ & 0.85 (20) & $7.0^{+0.3}_{-0.3}$ & 0.950 (957) & $9.5^{+1.3}_{-1.7}$ & 0.936 (953) \\
\hline
&&&&&&& \\

\end{tabular}
\end{center}

\parbox {7in}
{ Notes: A comparison of the temperature results obtained with the Single-phase (SP) and 
MultiPhase (MP) models. SP results from ASCA are for spectral Model B. MP results are for spectral Model
D. }
\end{table*}

One of the most important results from this paper is the marked 
difference in the temperatures of the clusters determined from the single-phase and multiphase
spectral models. These results are summarized in Table 13.  The single-phase temperatures 
consistently (and significantly) underestimate the multiphase results. 
For Zwicky 3146 and Abell 1835, the ASCA single-phase results underestimate the
multiphase temperatures by $\sim 10$ and 25 per cent,  respectively. 
With the ROSAT data, the discrepancy in much more severe, with the PSPC values
underestimating the ASCA multiphase results by $\sim 3.4 $ keV (50 per cent) for Zwicky 3146, and 
$\sim 6$ keV (60 per cent) for Abell 1835. Note, however, that in all cases the 
reduced $\chi^2$ values indicate statistically acceptable fits. 

The multiphase $kT$ results (Models C,D) should approximate the 
true mass-weighted temperatures in the clusters 
(Thomas \etal 1987; Waxman \& Miralda-Escude 1995; Allen \etal  
1996). The single-phase results, however, are simply emission/detector-weighted average
values for the integrated cluster emission. 
Since the X-ray emissivity of cluster gas rises with increasing density (decreasing temperature), 
the presence of a large cooling flow
naturally leads to a decrease in the emission-weighted temperature of a cluster.
The effects on the emission-weighted $kT$ are most dramatic 
in the $0.1-2.4$ keV band of the PSPC,  where 
the emission from cooler gas phases dominates the 
detected flux. However, the (comparatively) poor spectral resolution and limited band-pass
of the PSPC, mean that the single-phase models can still provide a statistically adequate description of 
the data. 
The PSPC data are unable to discern the need for multi-temperature components
(although the imaging data clearly require them). 

These results imply that caution should be applied in the interpretation of temperatures
determined with simple, single-phase models and, in particular, those determined 
from ROSAT data. 
With ASCA data the single-phase results should be more reliable, although significant 
discrepancies can still arise (as in the case of Abell 1835). 

The presence of a range of density and temperature phases is clearly established by the 
data for cooling flow clusters. However, it should not assumed that the absence of a cooling flow
implies that a single-phase modelling of the ICM is appropriate.
The existence of cooling flows with distributed mass deposition requires significant 
inhomogeneity (a density/temperature spread of $\sim$ a factor 2) in the ambient cluster
gas before the cooling flow forms (Nulsen 1986; Thomas, Fabian \& Nulsen 1987). The best data for clusters 
are consistent with such a range of inhomogeneity (Allen \etal 1992b). The absence of cooling flows 
in some nearby, luminous clusters such as the Coma cluster is usually attributed to merger events 
having disrupted the cluster cores and having re-heated and redistributed the cooling gas throughout the 
cluster. In such circumstances it seems unlikely that
the merger will completely homogenize the gas and, therefore, that a single-phase 
model will provide an exact measure of the mass-weighted cluster temperature.

\subsection{A comparison with lensing masses}

\begin{figure*}
\vskip 12.5cm
\caption{ The Hale 5m U band image of Abell 1835. 
Arc `A' is indicated to the South East of the CCG.
}
\end{figure*}

\begin{figure}
\centerline{\hspace{3.2cm}\psfig{figure=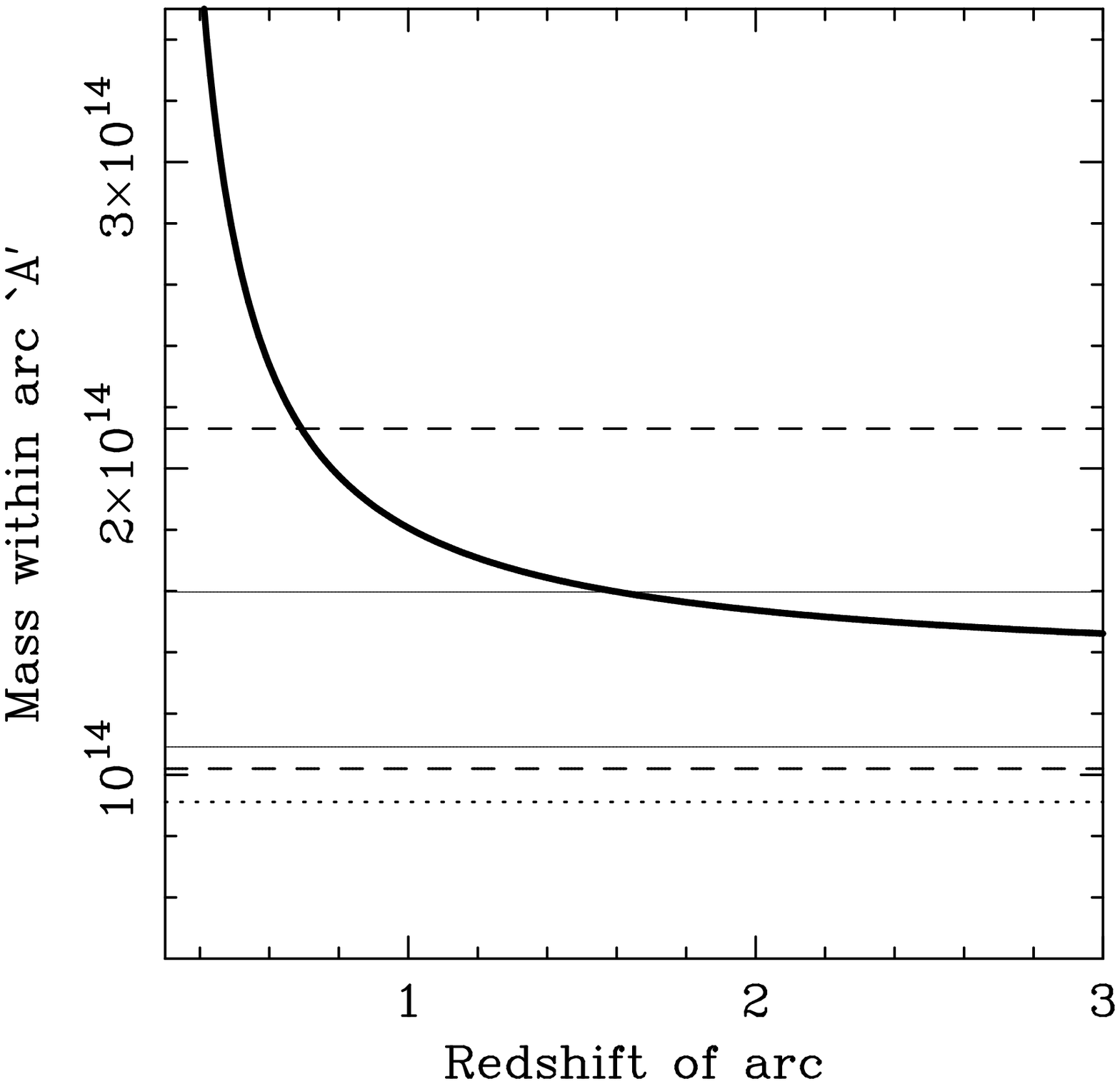,width=0.7\textwidth,angle=270}}
\caption{A comparison of the X-ray and lensing mass estimates for arc `A'. The bold 
curve shows the mass within the arc determined from the standard lensing formula
(for a spherical mass distribution) as a function of the redshift of the arc.
The solid horizontal lines show the X-ray constraints on the mass within this
radius determined with spectral Model D ($7.8 < kT < 10.8$ keV).  The 
dashed lines show the constraints for spectral Model C
($7.5 < kT < 14.4$ keV). The lensing and multiphase X-ray results together 
imply $z_{\rm arc} > 1.6$ for Model D,  and $z_{\rm arc} > 0.7$ for Model C.
The lower dotted line shows the mass within arc `A' 
implied by the single-phase spectral results ($M_{\rm proj} \sim 9.1 \times 10^{13}$ \Msun for
$kT = 7.0$ keV). The single-phase X-ray results are inconsistent with the lensing data.}
\end{figure}

In Fig. 13 we show the 
Palomar U band image of the central 2.5 arcmin$^2$ of Abell 1835. The CCG is the bright 
source in the centre of the field. The image shows a number of 
distorted features 
(arcs, arclets and image pairs) attributable to gravitational lensing by
the cluster.
In particular, we observe a bright, elongated arc 
(`A' in Fig. 13) at a radius of 30.3 arcsec from the centre of the CCG,  
along a PA of 133 degree.  The arc has a length of $\sim
16$ arcsec, is extended along a  PA of 221 degree, and exhibits reflection
symmetry about the point  $14^{\rm h}01^{\rm m}03.7{\rm s}$,
$02^{\circ}52'21''$. 

For a simple, circular mass distribution the projected mass within the 
tangential critical radius, $r_{\rm ct}$, is given by

\begin{equation}
M_{\rm proj}(r_{\rm ct}) ~ =
\frac{c^2 }{4 G} \left( \frac{D_{\rm arc}} {D_{\rm clus} D_{{\rm
arc-clus}}}
\right) ~ r_{\rm ct}^2
\end{equation}

where $r_{\rm ct}$ can be approximated by the arc radius 
($r_{\rm arc} = 150$ kpc) and 
 $D_{\rm clus}$,  $D_{\rm arc}$ and $D_{\rm arc-clus}$ are respectively 
the angular
diameter distances from the observer to the cluster, the observer to the
lensed object, and the  cluster to the lensed object. 
In Fig. 14 we show the mass within arc `A' as a function of the redshift of 
the arc, calculated with the above formula. 
Also shown are the X-ray constraints on 
the projected mass within this radius [$1.0 \times
10^{14}$ \Msun $< M_{\rm proj} < 2.1 \times 10^{14}$ \Msun~from the
multiphase analysis using spectral Model
C, and  $1.1 \times 10^{14}$ \Msun $ < M_{\rm proj} < 1.6 \times 10^{14}$ 
\Msun~using spectral Model D].  
Combining the X-ray and lensing mass results we are able 
constrain the redshift of arc `A' to be $> 0.7$.  
It is also important to note that if the single-phase 
(Model B) X-ray temperature results for Abell 1835 were (wrongly) used in
place of the multiphase values, 
no consistent solution for the X-ray and lensing masses would be possible. 

The intrinsic ellipticity of the lensing potential may lead to a 
slight ($\approxlt 20$ per cent) overestimate of the lensing mass 
determined with the circular mass model (Bartlemann 1995).
However, the effects of ellipticity also lead to a slight overestimate  
of the X-ray mass within this aperture and, to first order, the conclusions 
on the redshift of the arc should not be dramatically affected. 
[The lensing properties of Abell 1835, and
those of a larger sample of X-ray luminous clusters observed with the Palomar
5m telescope, are discussed further by Edge \etal (1996). ]

Smail \etal (1995) report results from a study of (weakly) gravitationally 
distorted images in the field of E1455+223, from which they derive a  projected
mass within 450 kpc of the cluster centre of $\sim 3.6 \times 10^{14}$ \Msun.
This mass exceeds the  X-ray determination of the projected mass within this
radius, $1.6 \times 10^{14}$ \Msun~(for an isothermal mass distribution
corresponding to a temperature of 5.4 keV), 
 by a factor $\sim 2$. [Note that the
determination of the X-ray mass assumes that the  cluster remains isothermal
and extends to 3Mpc. However, extrapolating the  mass profile to 5 Mpc
increases the projected mass within 450 kpc of the cluster centre by only $\sim
2$ per cent.] Using a potential consistent with the upper-limit to the 
X-ray temperature of  7.6 keV (\ie $\sigma = 900$
\kmps, $r_c = 40$ kpc) we still determine a  projected mass within 450 kpc of
only $2.5 \times 10^{14}$ \Msun. 

The lensing result on the cluster mass  for E1455+223 appears high. 
E1455+223 is a regular, relaxed cluster with a large cooling
flow, and  a $2-10$ keV X-ray luminosity of $1.3 \times 10^{45}$ \ergps. The
ASCA constraints on the X-ray temperature ($4.3-7.6$ keV)  are
consistent with results for other nearby, cooling-flow clusters of similar
X-ray luminosity (\eg Abell 1795;  Edge \etal 1990, Fabian \etal 1996), which
lends support to the X-ray mass determination. The velocity dispersion
of $\sigma = 660-900$ \kmps (Table 11) implied by the X-ray
data is also in good agreement with optical observations 
($\sigma \sim 700$ \kmps; Mason \etal 1981, Smail \etal 1995). 

A cluster of exceptional X-ray luminosity 
and temperature is required to provide a projected mass 
within 450 kpc consistent with the Smail \etal (1995) result for E1455+223. 
Abell 1835, discussed in this paper, provides a mass
of $2.5-5.3 \times 10^{14}$ \Msun. Similarly, the 
exceptionally X-ray
luminous cooling-flow  cluster 
PKS0745-191 ($L_X = 2.8 \times 10^{45}$ \ergps),  for
which  Allen \etal (1996) present a self-consistent determination of the
 mass distribution from X-ray and gravitational lensing data, provides 
a projected mass within 450 kpc of only $\sim 
3.7 \times 10^{14}$ \Msun. Given the X-ray luminosity of E1455+223 (which is
a factor 2-3 less than PKS0745-191 or Abell 1835), the 
X-ray mass measurement for the cluster seems reasonable 
and the lensing mass high. The result of Smail \etal (1995) may imply an unusual redshift
distribution for the weakly distorted sources, or a projected mass
distribution that deviates significantly from the simple isothermal mass
model used (perhaps due to 
some line of sight mass enhancement from material external to the X-ray
luminous part of the cluster).

\subsection{Optical and X-ray properties }

CCGs in cooling flows frequently exhibit characteristic 
low-ionization, optical emission-line 
spectra (\eg Johnstone, Fabian \& Nulsen 1987; Heckman \etal 
1989; Crawford \& Fabian 1992; Allen \etal 1995). The optical 
(H$\alpha\lambda6563$) line luminosity correlates with 
the excess UV/blue continuum luminosity 
(the excess with respect to the UV/blue emission 
expected from a normal gE/cD galaxy; 
Johnstone, Fabian \& Nulsen 1987; McNamara \& O'Connell 1989; Allen \etal 1992a; 
Crawford \etal 1995; Allen 1995). 
Typically, both the emission lines and the excess UV/blue continua 
are extended across the 
central $10-20$ kpc of the clusters. 

The clusters discussed in this paper contain three of the largest cooling 
flows known. They also host three of the most optically line-luminous 
(and UV/blue luminous) CCGs. Only the CCG of the massive cooling flow cluster 
PKS0745-191 exhibits a
comparable optical line luminosity (Allen \etal 1996). 
Although no simple correlation
between ${\dot M}$, $t_{\rm cool}$ and the optical line 
luminosity (and therefore
UV/blue continuum luminosity) exists, the data presented here 
confirm a tendency for the most optically-line-luminous CCGs 
to be found in the largest cooling flows (\eg Allen \etal 1995).   

The UV/blue continua in Abell 1835 and Zwicky 3146 appear dominated by 
emission from 
hot, massive stars (Allen 1995). These young stellar populations may also
provide the bulk of the ionizing continuum emission responsible for 
the observed optical emission lines. 
The exceptionally high mass deposition rates from the cooling flows in 
the clusters can naturally provide the 
large reservoirs of cooled material necessary to fuel the
observed (very high) star formation rates (Allen 1995). 
The star formation may also account for some of
the dust in the CCGs. 

The excellent alignments between the (optical) CCG and (X-ray) cluster isophotes 
are consistent with the results for other massive cooling flows at
intermediate ($z \sim 0.15$) redshifts (Allen \etal 1995).  These results again 
reveal the unique and intimate link between CCGs and their host clusters.

\section{CONCLUSIONS} 

We have presented detailed results on the X-ray properties of 
Zwicky 3146,  Abell 1835
and E1455+223, three of the most distant, X-ray luminous clusters known.  
We have shown that  these clusters contain the 
three largest cooling flows known, with mass deposition rates of $\sim 1400, 2300$
and 1500 \Msunpyr, respectively. We have presented mass models for the clusters and 
have highlighted the need for multiphase analyses to consistently explain 
the spectral and imaging X-ray data for these systems. 
The inappropriate use of single-phase models leads to significant underestimates 
of the cluster temperatures and masses.
For Abell 1835 it was shown that a mass distribution 
that can consistently explain both the X-ray and gravitational lensing data 
for the cluster can only be formed 
when multiphase X-ray spectral models are used. 
We have discussed the relationship between intrinsic X-ray absorption 
and optical reddening in the clusters. These results suggest that the X-ray
absorbing material  frequently observed in the X-ray spectra of cooling
flows is dusty.

\section*{Acknowledgments}

SWA, ACF and ACE thank the Royal Society for support. 
We thank I. Smail for communicating the results on the galaxy 
photometry in Abell 1835 and E1455+223.

\end{document}